\documentclass
[aps,twocolumn,floatfix,superscriptaddress,nofootinbib,citeautoscript,longbibliography]{revtex4-1}%
\usepackage{graphicx}
\usepackage{amsmath, mathtools}
\usepackage{amsfonts}
\usepackage{amssymb}
\usepackage{placeins}
\usepackage{helvet}
\usepackage{dcolumn}
\usepackage{tabularx,multirow}
\usepackage[english]{babel}
\usepackage[utf8]{inputenc}
\usepackage{epstopdf}
\usepackage{epsfig,color,textcase}
\usepackage[pdfpagemode={UseOutlines},bookmarks=true,bookmarksopen=true,
bookmarksopenlevel=0,bookmarksnumbered=true,hypertexnames=false,
colorlinks,linkcolor={blue},citecolor={blue},urlcolor={blue},
pdfstartview={FitV},unicode,breaklinks=true]{hyperref}
\usepackage{xspace}

\DeclareMathOperator\erfc{erfc}
\let\tn\textnormal

\begin{document}

\title{Microscopic theory of magnetic detwinning in \\ iron-based superconductors with large-spin rare earths}
\author{Jannis Maiwald}
\affiliation{Experimentalphysik VI, Universit\"at Augsburg, Universit\"atstra{\ss }e 1,
86135 Augsburg, Germany}
\author{I. I. Mazin}
\affiliation{Code 6393, Naval Research Laboratory, Washington, DC 20375, USA}
\author{Philipp Gegenwart}
\affiliation{Experimentalphysik VI, Universit\"at Augsburg, Universit\"atstra{\ss }e 1,
86135 Augsburg, Germany}
\date{\today}

\begin{abstract}
Detwinning of magnetic (nematic) domains in Fe-based superconductors has so
far only been obtained through mechanical straining, which considerably
perturbs the ground state of these materials. The recently discovered
non-mechanical detwinning in EuFe$_{2}$As$_{2}$ by ultra-low magnetic fields
offers an entirely different, non-perturbing way to achieve the same goal.
However, this way seemed risky due to the lack of a microscopic understanding of
the magnetically driven detwinning. Specifically, the following issues
remained unexplained: (i) ultra-low value of the first detwinning field of $\sim
0.1$\,T, two orders of magnitude below that of BaFe$_{2}$As$_{2}$, (ii)
reversal of the preferential domain orientation at $\sim1$\,T, and restoration
of the low-field orientation above 10--15 T. In this paper, we present, using
published as well as newly measured data, a full theory that quantitatively
explains all the observations. The key ingredient of this theory is a
biquadratic coupling between Fe and Eu spins, analogous to the Fe-Fe
biquadratic coupling that drives the nematic transition in this family of materials.

\end{abstract}

\maketitle

\email{mazin@nrl.navy.mil}

\textbf{Introduction }One of the most admirable experimental feats in studies
of Fe-based superconductors (FeBS) is the mechanical detwinning of the
low-temperature phases of the parent compounds in the 122
families\cite{Ian,Ruslan}. This allowed impressive insight into the physics of
spin-driven nematicity, a phenomenon that arguably rivals the
superconductivity itself in these materials. In this connection, one of the
most intriguing and unexpected findings was that this nematic physics is
ensured by a sizable biquadratic magnetic interaction, something unheard of in
localized magnetic moment systems, and never investigated in itinerant
magnetic metals. This phenomenon was first discovered
computationally\cite{Yaresko} and later shown to provide the only physically
meaningful description of spin dynamics in FeBS\cite{Kirill}. There is growing
evidence that it is not limited to FeBS, but occurs also in other itinerant
systems\cite{BTSO}.

\begin{figure*}
\includegraphics[width=\textwidth]{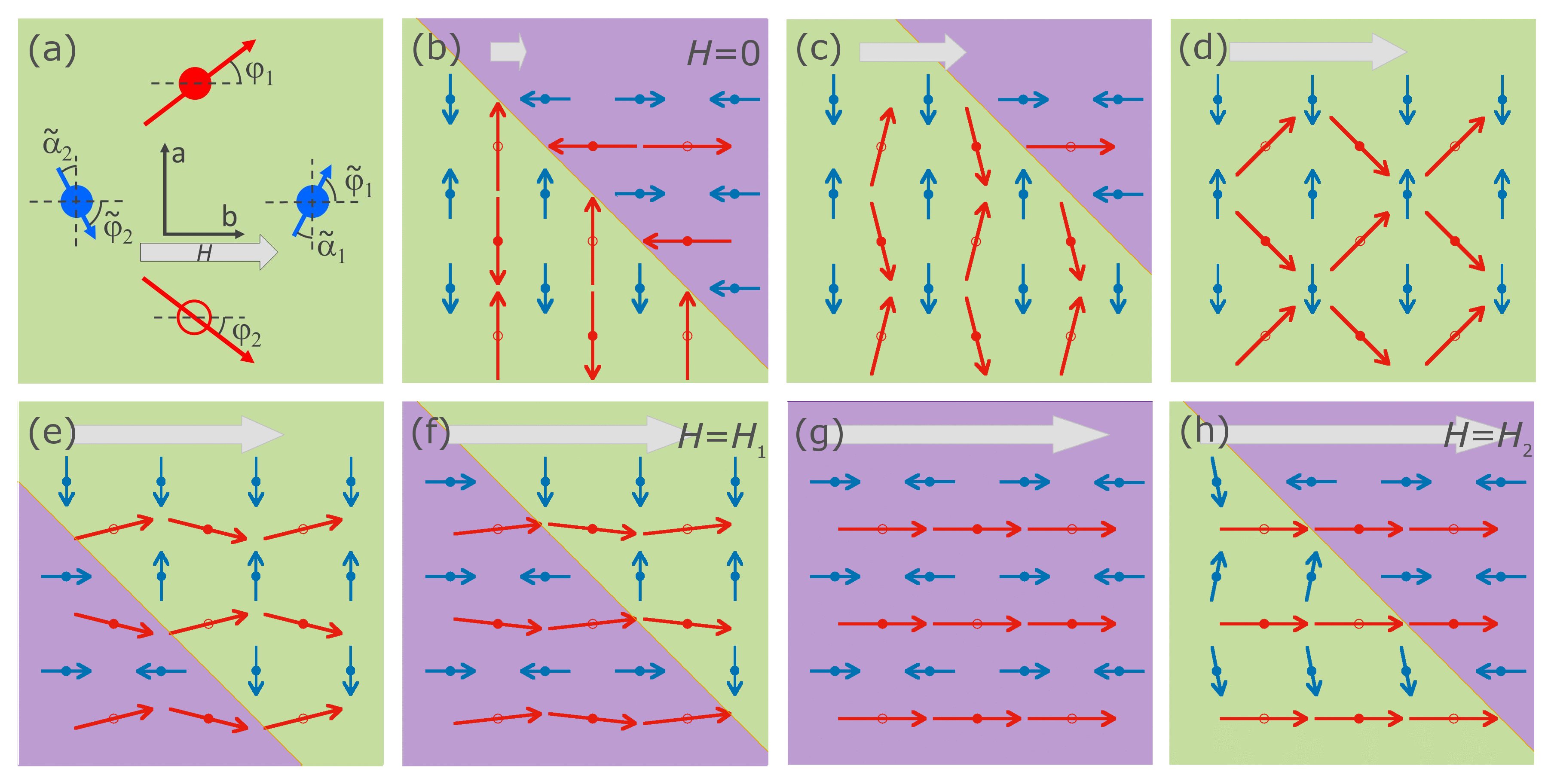}
\caption{Illustration of the low-temperature kinetic detwinning process of EuFe$_2$As$_2$ for
 an (from left to right) increasing external magnetic field $H\parallel [110]_\text{T}$ (gray arrow). Fe atoms and
 spins are shown in blue. Open circles indicate Eu atoms and spins (red) in the next layer. (a) Definition of the
spin angles. (b) Initial twin population (b-twin in green, a-twin in purple) at $H=0$. (c) The b-twins grow
 with
increasing field until (d) the system is completely detwinned with $b\parallel H$. (e) Above $H_0$ the 
reorientation to a-twins sets in, (f) which is largest at $H_1$ and (g) completes above, with $a\parallel
H$. (h) Around $H_3$ the final reorientation back to b-twin occurs.}
\label{fig:dynamics}%
\end{figure*}

Mechanical straining is not the only way to detwin FeBS. It was shown that a
static magnetic field of $\sim15$\thinspace T leads to partial
detwinning\cite{Ian1} and pulsed fields of $\sim30$\thinspace T to nearly
complete detwinning\cite{Ian2}. Later we will analyze these facts in more
detail, but at the moment we emphasize that these are relatively large fields,
even though the in-plane magnetic anisotropy energy of the FeAs planes was
experimentally shown to be of the order of 0.5\thinspace meV and, therefore,
sizable compared to e.g. elemental Fe, where it is only a few $\mu$eV. Against
this background, it came as a complete surprise when it was
discovered\cite{Eu1,Jannis} that substituting Ba with Eu lowers the field
needed for full detwinning by two orders of magnitude. In principle, this
magnetic detwinning allows for a virtually non-intrusive (the energy scale
associated with this field is less than 20\thinspace mK) investigation of the
physics of the nematic state. However, this seemingly exciting opportunity was
met with limited enthusiasm for the simple reason that no plausible
microscopic explanation could be found for the detwinning itself, given the
minuscule amplitude of the required field. Even more striking was the
discovery that by increasing the magnetic field gradually one can switch the
sign of detwinning \textit{twice}: initially, twin domains orient
 in such a way  that the Fe-Fe ferromagnetic bonds along the crystallographic $b$ 
are parallel to the applied field (we call this
the b-twin, see Fig.~\ref{fig:dynamics}a). This process is essentially complete at $H_{0}\sim0.5$ T. Then, at
$H_{1}\sim1$ T domains spontaneously rotate in-plane by 90$^{o},$ and at $H_{2}\sim10$
T start to turn back to the detwinned state that was initially generated at
$H\lesssim0.5$ T, see also Fig.~\ref{fig:dynamics}(b-h). With such a complex phase diagram, and no theoretical
understanding of the underlying phenomena, it is indeed worrisome to embark on
systematic studies of nematicity with the risk that unknown magnetic physics may
affect the findings. The goal of this paper is to remedy this situation and
present a full and quantitative theory explaining all the above observations.
It appears that magnetically induced detwinning is intimately related to the
nature of nematicity itself, namely, it is also driven by a sizable
biquadratic interaction, which, in turn, is the consequence of the Janusian
itinerant-localized nature of the FeBS. 

\textbf{Formalism }We start with the (simpler) case of BaFe$_{2}$As$_{2}.
$ The minimal approximation is the nearest-neighbor (n.n) Heisenberg model
with single site anisotropy:%
\begin{equation}
\mathcal{H}_\text{Fe}=-\tilde{D}\sum_{i}f_{i,a}^{2}+\tilde{J}\sum_{\left\langle
ij\right\rangle }\mathbf{f}_{i}\cdot\mathbf{f}_{j}-\tilde{M}\sum_{i}%
\mathbf{f}_{i}\cdot\mathbf{H}\label{Hfe}%
\end{equation}
where $i,j$ label Fe sites, $\mathbf{f}_{i}$ is the unit vector directed along
the Fe magnetic moment at site $i,$ $\tilde{M}$ is its absolute value, and the
summation is over all inequivalent n.n bonds\cite{note}. $\mathbf{H}$ is the
external field in the corresponding energy units. Here and below we use tildes
over symbols for Fe-related parameters.

It is known experimentally that the Fe moments lie within the $ab$ plane and
are oriented along the longer $a$-axis ($i.e.$, $\tilde{D}>0)$. Neutron
scattering provides estimates for the parameters as $\tilde{J}\sim
30$\thinspace meV, $\tilde{D}\approx0.25$\thinspace meV \cite{neutrons}.
Minimizing Eq.~\eqref{Hfe} with respect to \textbf{f, }for $\mathbf{H||}b$ (no
linear susceptibility appears for $\mathbf{H||}a)$ we observe that $H$
generates a canting of the Fe spins away from the $a$ axis by an angle
$\tilde{\alpha}=\sin^{-1}(\tilde{M}H/4\tilde{J})$, which corresponds to an energy gain of
$E=(\tilde{M}H)^{2}/4\tilde{J}$ per formula unit (f.u). There are two ways how
the system may take advantage of this energy gain, even if the field is along
$a$. First, all Fe spins may rearrange (abruptly) from being aligned along $a$ to being
predominantly aligned along $b$. This process is called spin-flop and occurs
when $\tilde{J}>\tilde{D}.$ Using $\tilde{M}\approx1\,\mu_{\text{B}} $ and the
above parameters, we estimate $\tilde{H}^{\text{flop}}=2\sqrt{2\tilde{D}%
\tilde{J}}/\tilde{M}\approx11\,\tn{meV}\approx135$\thinspace T. Thus, the
spin-flop never occurs at typically lab-accessible fields. The other way is to
switch an entire \textquotedblleft a-twin\textquotedblright\ (\textit{i.e.},
a domain with antiferromagnetic bonds along the crystallographic $a$ in field direction) to a \textquotedblleft
b-twin\textquotedblright (in other words, to rotate the crystal structure by 90$^{o}$
 around $c$, keeping the moments aligned along the magnetic easy $a$ axis. This process
is associated with an unknown energy
barrier $\Delta.$ Given that in the field of $\sim15$\thinspace T the Stanford
group has observed partial\cite{Ian1}, and in $\sim30$\thinspace T nearly full
detwinning\cite{Ian2}, we deduce $\Delta\sim E=(\tilde{M}H)^{2}/4\tilde
{J}\approx0.025$\thinspace meV/f.u.

One can verify these deductions against mechanical detwinning. The latter is
an indirect process wherein it is difficult to access the microscopic strain
and stress. Reported values for the latter differ from $\sigma=$
6--20\thinspace MPa\cite{Jiang13,Makar}. Assuming $\sigma=10$\thinspace MPa
and taking the elastic modulus to be 10\,GPa\cite{Shein09,Yoshizawa12}, we derive a
strain of $\varepsilon\approx$ 0.1\%. Finally, using the calculated dependence
of the total energy on the microscopic strain\cite{Dresden}, we find that this
strain corresponds to $\Delta\approx$ 0.01 \thinspace meV/f.u, which is in
agreement with our magnetic estimate.

Yet another estimate can be obtained by considering the stress $\sigma$ on the
unit cell during mechanical detwinning and calculate the energy associated
with the displacement $\delta$ from $a\parallel\sigma$ to $b\parallel\sigma$.
Using the reported lattice parameters of EuFe$_2$As$_2$\cite{Xiao09}, we arrive at
$\delta=$ 1.6$\times10^{-12} $ m. Assuming $\sigma=6$\thinspace MPa, this
leads also to an energy of $\Delta\approx0.01$\thinspace meV/f.u.

Given that the detwinning energy, according to this theory, depends
quadratically on the magnetic moment, and inversely on the exchange constant,
one may naively assume that the same mechanism will be operative in EuFe$_{2}%
$As$_{2},$ given that the ordering temperature in the Eu sublattice is much
smaller, $T_{\tn{N}}=19$\thinspace K, and the moment much larger,
$M=7\,\mu_{\tn{B}},$ than for the Fe sublattice. Such a phenomenology was
adapted in Ref. \cite{Jannis} in order to parametrize the observed effect.
However, it is easy to see that it is microscopically untenable. Indeed, while
it is possible and reasonable to write down interactions $inside$ the Eu
sublattice in Heisenberg form,
\begin{equation}
\mathcal{H}_\text{Eu}=J_{\parallel}\sum_{\left\langle ij\right\rangle }%
\mathbf{e}_{i}\cdot\mathbf{e}_{j}+J_{\perp}\sum_{i}\mathbf{e}_{i}%
\cdot\mathbf{e}_{i^{+}}-M\sum_{i}\mathbf{e}_{i}\cdot\mathbf{H}\label{HEu}%
\end{equation}
where $i,j$ label Eu sites, $i^{+}$ the opposing Eu site in the next layer,
$\mathbf{e}_{i}$ the unit vectors directed along the Eu magnetic moment $M$ at
site $i$, and the ferromagnetic $J_{\parallel}<0$ and antiferromagnetic
$J_{\perp}>0$ constants determine the in-plane and interplanar ordering,
respectively, it is not possible to describe the interaction between the Fe
and Eu subsystems in the same manner, for the simple reason that the
Heisenberg exchange field induced by the Fe planes on the Eu sites is zero by
symmetry. In fact, any bilinear coupling
between the Fe stripes and in-plane Eu spins is zero by symmetry, including Heisenberg, 
Dzyaloshinky-Moria and dipole interaction (in the last case the field induced by the Fe
planes on the Eu sites is non-zero, but is directed strictly along $z$, see supplement). Note that we 
did not include any
single-site anisotropy in Eq.~\eqref{HEu}, because Eu adopts a valence state
of $+2$ in this system. Due to the closed $f-$shell, with 7 electrons in the
spin-majority channel, Eu$^{2+}$ has zero angular momentum and negligible
magnetocrystalline anisotropy. This is confirmed by first principles
calculations, presented in the supplement to this article. Without an
interaction between Fe and Eu, there is no physical mechanism by which the Eu
spin dynamics may affect the detwinning.

Another intriguing problem, possibly related to this one, is the fact that
even the basic magnetic properties of EuFe$_{2}$As$_{2}$ cannot be
explained within a simple Heisenberg model. Indeed, the magnetic susceptibility
of EuFe$_{2}$As$_{2}$ above $T_{N}$ is dominated by the Eu spins and well
described by a nearly isotropic Curie-Weiss law, in accordance with the
previous paragraph.
Eq.~\eqref{HEu} suggests that $k_{\text{B}}T_{\text{CW}}= \frac{3}{7} (4J_{\parallel}-2J_{\perp
})$ (the quantum prefactor $3M/(M+2)$ would have been 1 if $M$ were $1\,\mu_\text{B}$, but for
$M=7\,\mu_\text{B}$ it becomes $3/7$)
The effective moment has been determined to be
$M=7\,\mu_{B}$ with $T_\text{CW}\approx-20$\thinspace K \cite{Jiang09}.
Thus the N\'{e}el temperature appears to be equal to
the mean-field transition temperature of the individual Eu planes. In other
words, each Eu plane orders magnetically at the mean-field temperature, not at
all suppressed by fluctuations, and immediately at the transition the
antiferromagnetic stacking of the individual planes along $c$ is acquired.

At this point, it is instructive to look at first principles calculations and
what they tell us about $J_{\perp}$ and $J_{\parallel}.$ The former appears to
be very small and decreases with the value of the Hubbard $U$ used on the Eu $f$
orbitals. This is not surprising, because it is set by the competition between
superexchange, proportional to $t_{\perp}^{2}/U,$ and Schrieffer-Wolff
driven\cite{DMS} double exchange, proportional to ($t_{\tn{Eu-Fe}}^{2}%
/U)^{2}N(0),$ with $N(0)$ the density of states, and $t_{\perp}$ and
$t_{\tn{Eu-Fe}}$ the effective Eu-Eu and Eu-Fe hopping across the planes. For
$U-J=0$ we get $J_{\perp}=0.14$\thinspace meV (1.6\thinspace K) and $J_{\parallel}=-2$\thinspace meV (23\thinspace K), while for
$U-J=5$\thinspace eV we find $J_{\perp}=0.26$\thinspace meV ($3$\thinspace K) and $J_{\parallel}=$-0.8\thinspace meV (9\thinspace K). The LDA+$U$ results, which we believe are
closer to reality, correspond to $T_{\tn{CW}}\sim-13$\thinspace K, and
$T_{\tn{N}}\approx4.27J_{\parallel}/[3.12+\log(J_{\parallel}/J_{\perp}%
)]=9$\thinspace K for the Heisenberg model\cite{Heisenberg2+1}. The ratio $T_{\tn{CW}}/T_{\tn{N}}$
is close to 1.4 rather than the experimental 1.13, and one can see that in order to reduce it to 1.13
one needs to increase the $J_\perp/J_\parallel$ ratio to $\sim 0.5$, a rather 
unrealistic 3-dimensionality.

The situation could be remedied if one were to assume a finite single-site anisotropy for Eu of
$D\approx 1$\,K, because in such a case one has to replace $J_\perp$ with 
$J_\text{eff}= J_\perp+D+\sqrt{D^2+2J_{\perp}D}$ \cite{Irkhin98,Irkhin99}, which leads to a sufficient increase
of the N\'{e}el temperature and consequently to $T_{\tn{CW}}/T_{\tn{N}}$ = 1.13.
But, as we have argued above, Eu$^{2+}$ has no single-site anisotropy. One of the results of our paper,
however, is that a biquadratic coupling $K$ 
between the Eu and Fe subsystems is operative in EuFe$_{2}$As$_{2}$, which plays the key role in its
magnetic detwinning. This interaction acts as an \textit{effective anisotropy} for the Eu subsystem,
which tries to imprint the orientation of the Fe spins on the Eu spins. The corresponding number
($8K$) that we extract from the experiment amounts to $D_\text{eff}\sim 0.6$ K.
Substituting this into the formulas above we get $T_{\tn{CW}}/T_{\tn{N}}\approx 1.18$, which is within the
error margin of the experimental ratio. 
Of course, given the model character of these calculation,
a discrepancy of $\approx 25$\% is not too alarming.
However, it is noteworthy that after adding the experimentally determined 
biquadratic term the discrepancy virtually disappears completely.

Note that, unlike correlated insulators, FeBS have a considerable biquadratic 
interaction within the Fe subsystem, which plays a
crucial role in their nematic behavior\cite{FeSe}. Whether this is a universal
property of correlated magnetic metals, which simply had not been given proper
attention before, or is unique for FeBS is unknown. With this in mind, we have
combined the Hamiltonians from Eqs.~\eqref{Hfe} and \eqref{HEu} in the
following way:%
\begin{align}
\mathcal{H}=&-\tilde{M}\sum_{\alpha}\mathbf{f}_{\alpha}\cdot
\mathbf{H}+\tilde{J}\sum_{\left\langle \alpha\beta\right\rangle }%
\mathbf{f}_{\alpha}\cdot\mathbf{f}_{\beta}-\tilde{D}\sum_{\alpha}f_{\alpha
,a}^{2}\nonumber\\
&-M\sum_{i}\mathbf{e}_{i}\cdot\mathbf{H}+J_{\perp}\sum_{i}\mathbf{e}%
_{i}\cdot\mathbf{e}_{i^{+}}\nonumber\\
&-K\sum_{\alpha,n}\left(  \mathbf{f}_{\alpha}\cdot\mathbf{e}_{\alpha
+n}\right)  ^{2},\label{eq:full_hamiltonian}%
\end{align}
where the Greek subscripts label the Fe sites, Latin the Eu sites, and the
last summation runs over all n.n Fe-Eu pairs. We have then estimated the biquadratic
Eu-Fe coupling term $K$ from LDA+U calculations (see SM), and obtained an
estimate of $K\sim0.4$ meV. Given the uncertainty in the calculations, this
should be taken as evidence that $K$ is not negligible; later we will
determine the actual amplitude of $K$ directly from the experiment.

In the Supplementary Information we present detailed derivations of the solutions
of Eq.~\eqref{eq:full_hamiltonian} for both possible orientations of the
external field with respect to the crystallographic axes, and for all relevant
field regimes. The discussion below omits less relevant parts of the full
theory, concentrating on rationalizing the actual experimental observations.

\textbf{Domain energetics} We start our discussion in the \textit{low field
}regime, that is $0<H\lesssim H_{1}\sim$1\,T, before considering higher fields. In
this regime the Fe single-site anisotropy dominates and Fe spins are always
oriented along the crystallographic $a$ axis, and so are, initially, Eu spins.
We will distinguish two cases: first, when $\mathbf{H}$ is applied
perpendicular to the initial orientation of Eu spins (b-twin) or,
second, parallel to those (a-twin). In this regime the former is always
lower in energy, since the latter has formally zero spin susceptibility. We
will characterize the orientation of the Eu and Fe spins by their respective angles $\varphi$ and $\tilde{\varphi}$ with respect to
the external field. Figure~\ref{fig:dynamics}(a) is drawn for a
 b-twin domain, i.e., $a\perp H$. Equation~\eqref{eq:full_hamiltonian} can then be rewritten in terms of these angles (per one
formula unit) as
\begin{widetext}
\begin{align}
\mathcal{E}=&-\tilde{M}H(\cos\tilde{\varphi}_1+\cos\tilde{\varphi_2})+2\tilde
{J}\cos(\tilde{\varphi_2}{+}\tilde{\varphi}_1)-\tilde{D}(\cos^{2}%
\tilde{\alpha}_1+\cos^{2}\tilde{\alpha_2})\nonumber\\
&  -MH(\cos\varphi_1+\cos\varphi_2)/2+J_{\perp}\cos{(}\varphi_2{+}\varphi_1)\nonumber\\
&  -2K[\cos^{2}(\varphi_2{-}\tilde{\varphi}_2)+\cos^{2}(\varphi_2{-}\tilde{\varphi
}_1) +\cos^{2}(\varphi_1{-}\tilde{\varphi}_2)+\cos^{2}(\varphi_1%
{-}\tilde{\varphi}_1)],
\end{align}
\end{widetext}
where $\alpha$'s are the Fe angles measured from the magnetic easy $a$-axis.

For the b-twin domains and low fields, i.e. $\tilde{\varphi}_{2}=\tilde{\varphi}%
_{1}=\pi/2,$ $\varphi_{1}=\varphi_{2}=\varphi,$ $\tilde{\alpha}_{2}%
=\tilde{\alpha}_{1}=0$ this yields
\begin{equation}
E_{b}=-MHp(2J_{\perp}+8K)p^{2}+E_{0},\label{eq:atwin}%
\end{equation}
where we have expressed everything in terms of $p=\cos\varphi$ and $E_{0}%
=-2\tilde{J}-2\tilde{D}-J_{\perp}-8K$. The equilibrium tilting angle and energy
are given by
\begin{equation}
p_{b}^{\text{min}}=\frac{MH}{4J_{\perp}+16K}%
\end{equation}
\begin{equation}
E_{b}^{\min}=-\frac{(MH)^{2}}{8(J_{\perp}+4K)}+E_{0}.
\end{equation}

The biquadratic term in the definition of $E_0$ is always trying to minimize the angle between the Fe
spins and Eu spins. Thus, in simplified terms (see supplement for details), if $\pi/4<\varphi<\pi/2,$ Fe spins prefer to orient
perpendicular to the field ($\tilde{\varphi}=\pi/2)$, and the b-twin domain
is always lower in energy. When $\varphi$ becomes smaller than $\pi/4,$ it
becomes more favorable to orient Fe spins parallel to the field $(\tilde
{\varphi}=0)$, and this is the first critical field $H_{1}$ at which the first
domain reorientation from b-twins to a-twins takes place (note that in this field regime only
$\tilde{\varphi}=\pi/2$, or, in case of the a-twin, 0 is allowed, any intermediate value of
$\tilde{\varphi}$ is severely punished by the Fe-Fe exchange and single-site anisotropy).

After this reorientation has occurred, the total energy is expressed as
\begin{equation}
E_{a}=-MHp+(2J_{\perp}-8K)p^{2}+E_{0}+8K%
\label{eq:E_a}
\end{equation}
and consequently
\begin{equation}
p_{a}^{\min}=\frac{MH}{4J_{\perp}-16K}%
\end{equation}
\begin{equation}
E_{a}^{\min}=-\frac{(MH)^{2}}{8(J_{\perp}-4K)}+E_{0}+8K.
\end{equation}
Thus the first reorientation field $H_{1}$ is defined by $E_{a}^{\min}%
=E_{b}^{\min},$ or%
\begin{equation}
H_{1}=\frac{2}{M}\sqrt{2(J_{\perp}^{2}-16K^{2})}%
\end{equation}

At the field $H_{a}^\text{sat}=4(J_{\perp}-4K)/M$ the angle $\varphi$ becomes $0,$
that is to say, Eu spins are perfectly aligned with the field. Further
increase of the field does not change the total energy (aside from the Zeeman
term $-MH)$, because from that point on the differential spin susceptibility
of the Eu subsystem becomes zero. However, while theoretically important, this
field does not manifest itself as a change of regime in domain dynamics.

$H_{a}^\text{sat}$ is too small to incur any Fe spin dynamics, but, in principle,
with further increase of $H$ one needs to include the spin susceptibility of
the Fe subsystem. The latter is zero as long as Fe spins are parallel to Eu
spins, satisfying the biquadratic coupling. Yet, in a sufficiently strong
field, i.e. at $H\gtrsim H_{2}$ a potential energy gain from allowing Fe spins
to screen the field outweighs the loss of the biquadratic interaction.
Mathematically, the latter, in this regime $H>H_{a}^\text{sat}$, is reduced to an
effective single-site anisotropy for the Fe subsystem, subtracting from the
actual anisotropy, and the transition in question is described by the same formulas as a typical spin-flop
transition. We can find the corresponding field value in the same way as one derives the spin-flop field in
textbooks: we need the energy gain from the Fe
spin-flop to overcome the energy loss due to noncollinearity (the loss occurs
both due to the Fe-Fe exchange $\tilde{J}$ and because the Fe site-anisotropy
is much larger than the biquadratic coupling).

In this case, the total energy (since now $\varphi=0,$ $\tilde{\alpha}%
=\pi/2-\tilde{\varphi})$ is
\begin{align*}
\tilde{E}_{b}= &  -2\tilde{M}H\tilde{p}+2(2\tilde{J}+\tilde{D}-4K)\tilde{p}^{2}\\
&  -MH+E_{0}+2J_{\perp}+8K,
\end{align*}
where we have now used $\tilde{p}=\cos\tilde{\varphi}$ and $E_{0}$. Minimizing with respect
to $\tilde{p}$ yields:
\[
\tilde{p}_b^\text{min}=\frac{\tilde{M}H}{2(2\tilde{J}+\tilde{D}-4K)},
\]
with the energy gain compared to Eq.~\eqref{eq:E_a} with $p=1$ being%
\[
dE=+\frac{(\tilde{M}H)^{2}}{2(2\tilde{J}+\tilde{D}-4K)}-8K.
\]
This expression changes sign at $H_{2}=4\sqrt{2\tilde
{J}K+\tilde{D}K-4K^{2}}/\tilde{M}\approx4\sqrt{2\tilde{J}K}$\textsc{/}$\tilde{M}.$ This
is the second critical field at which the reorientation back to the b-twin domains is initiated.

The domain dynamics, with the initial detwinning to b-twins at $H_0$, first reorientation to a-twins at $H_1$,
and second reorientation back to b-twins at $H_2$ is illustrated in Fig.~\ref{fig:dynamics}(b-h) and also in a supplemental movie (\textcolor[rgb]{1,0,0}{url}).
There, we depict how Eu and Fe spins (and the structural axes follow the
latter) rotate in an external field. One can see that, despite the physical
simplicity, the actual dynamics are rather complicated.

\textbf{Determination of the Coupling Constants} While the $H_{0,1,2}$ defined
above directly manifest themselves in the experiment, the actual expression
for the energy difference between the two types of domains (which is needed to
describe domain dynamics, as opposed to the thermodynamic equilibrium) is
a complicated piecewise function of the field. The full derivation can be
found in the Supplement, where explicit expressions for all critical fields
are obtained. In the relevant field range for the experiments discussed here
this function is given by%
\begin{equation}
dE=\left\{
\begin{array}
[c]{ll}%
\frac{M^{2}H^{2}}{8(J_{\perp}+4K)}, & \{0,H_{a}^{\text{flop}}\}\\
K(8-\frac{M^{2}H^{2}}{J_{\perp}^{2}-16K^{2}}), & \{H_{a}^\text{flop},H_{a}^{\text{sat}%
}\}\\
-MH+2J_{\perp}+\frac{M^{2}H^{2}}{8(J_{\perp}+4K)}, & \{H_{a}^{\text{sat}%
},H_{b}^{\text{sat}}\}\\
-8K+\frac{\tilde{M}^{2}H^{2}}{2(2\tilde{J}+\tilde{D}-4K)}, & \{H_{b}%
^{\text{sat}},H_{2}\}.
\end{array}
\right. \label{dEall}%
\end{equation} In the second case $dE>0$ changes to $dE<0$ at $H_1$, which lies 
between $H_{a}^\text{flop}$ and $H_{a}^\text{sat}$. The fields 
$H_{1},H_{a}^{\text{sat}}$ and $H_{2}$ have been derived above. The 
transformations associated with $H_{a}^{\text{flop}}$ and $H_{b}^{\text{sat}}$ occur in 
\textquotedblleft wrong\textquotedblright, or minority domains, which are 
thermodynamically unstable, but occur kinetically. Their expressions, as 
derived in the Supplement, are $H_{b}^{\text{sat}}=4(J_{\perp}+4K)/M$ and 
$H_{a}^{\text{flop}}=8/M\sqrt{K(J_{\perp}-4K)}.$

The coupling constants $J_{\perp}$ and $K$ can be determined from experiment.
Magnetization, magnetostriction, neutron and magneto-transport measurements,
for instance, all can be used to estimate the domain population ratio. In the
following we will determine the coupling constants using new magnetization
data\cite{Jannis}, and then use them to calculate this ratio as a function of 
field in order to compare it to the experiment.

\begin{figure}
\includegraphics[width=0.5\textwidth]{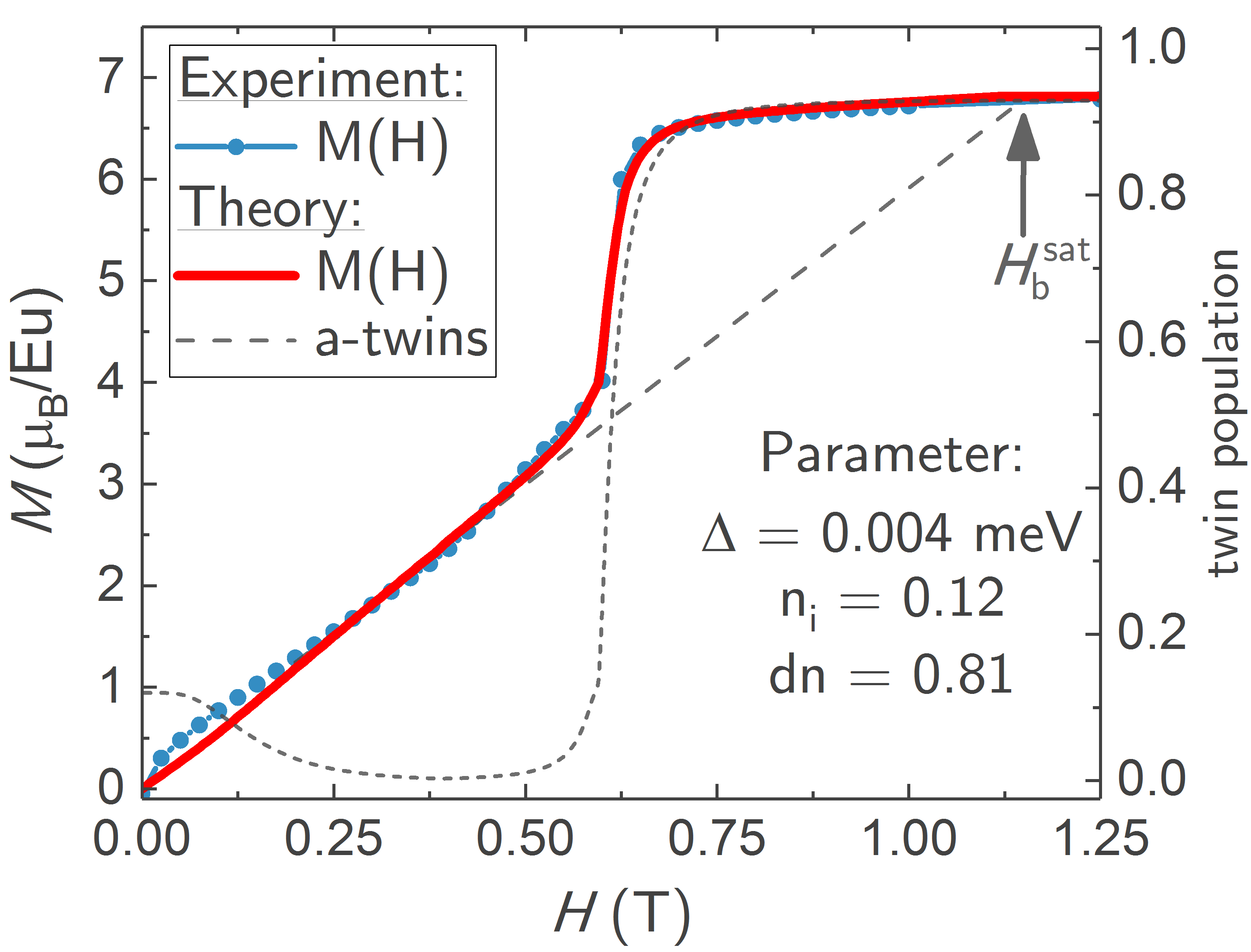}\caption{Magnetization
at ($T=5K$) as a function of decreasing magnetic field applied along the
$[110]_{\text{T}}$ direction (blue symbols) remeasured from Zapf et
al.\cite{Jannis}. The solid line (red) represents our theoretical prediction
using the determined coupling constants. The short dashed line depicts the
a-twin domain population $n$ derived from a fit to the corresponding
magneto-resistance data, similar to Fig.~\ref{fig:MS}, yielding:
$\Delta=0.004$\thinspace meV, $n_{i}=0.12$ and $dn=0.81$ (see Supplement). 
The long dashed line depicts the extrapolation of $H_b^\text{sat}$, as discussed in the text.}%
\label{fig:magnetization}%
\end{figure}

\begin{figure}
\includegraphics[width=0.5\textwidth]{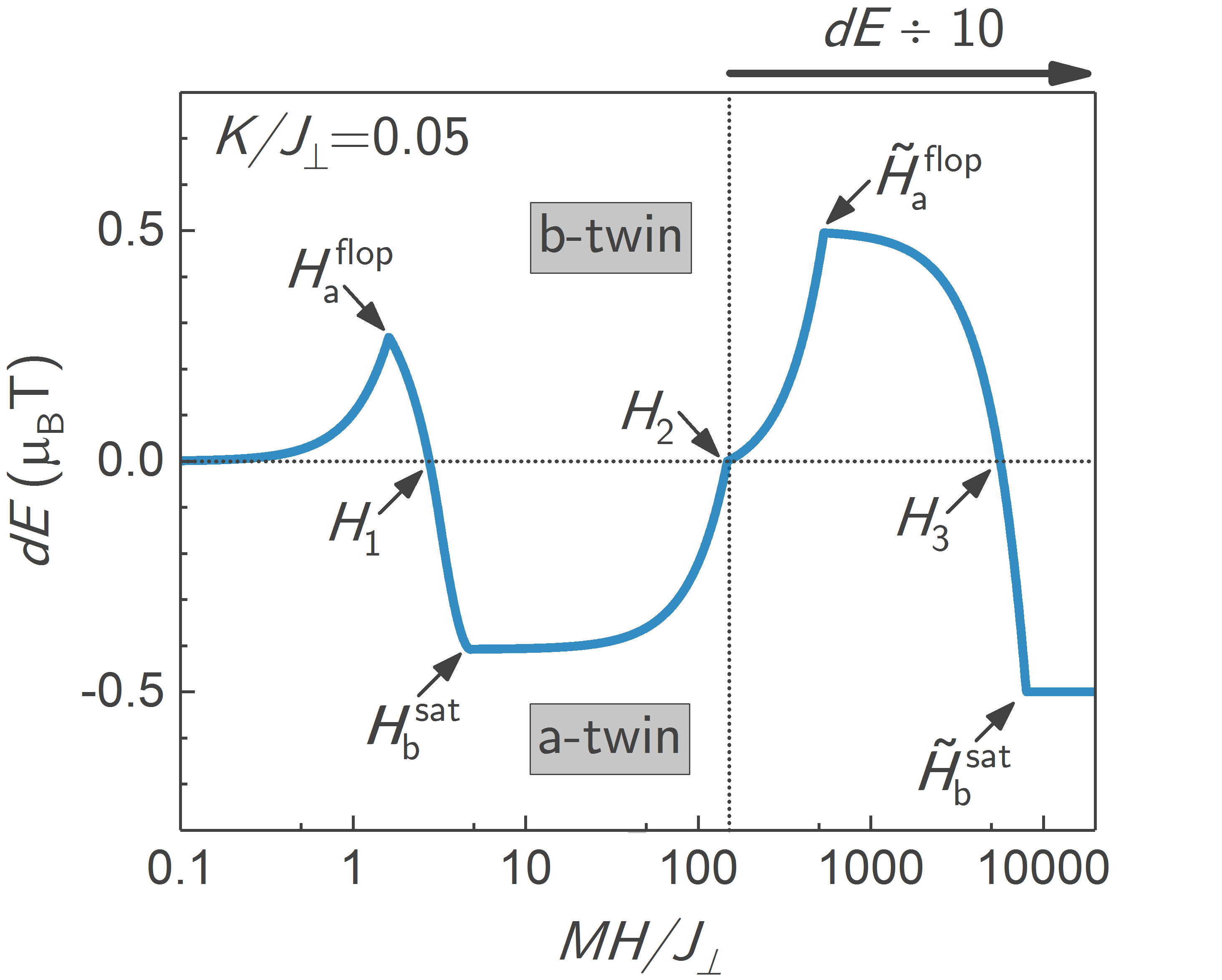}\caption{Calculated domain
energy difference $dE$ as a function of $MH/J_{\perp}$ for of a characteristic ratio of $K/J_{\perp}%
=0.05$. For better visibility, values above ${H}_{2}$ (vertical dashed line) are rescaled by a
factor of 10. Positive values and negative values correspond to the b-twin and a-twin domain, respectively. The horizontal dashed line indicates $dE=0$.}%
\label{fig:dE}%
\end{figure}

When measured along the tetragonal $[110]_{\text{T}}$ direction, a roughly linear
magnetization $M(H)$ was reported in Ref. \cite{Jannis}, interrupted only by
two pronounced jumps around 0.1\thinspace T and 0.6\thinspace T. Saturation
sets in above 0.9\thinspace T with $M\approx5\,\mu_{\text{B}}$. Due to the
uncharacteristically small saturated moment, which does not agree well with
previously published data\cite{Jiang09} and the theoretically expected value
of $M=7\,\mu_{\text{B}}$, we have re-measured the magnetization of a
EuFe$_{2}$As$_{2}$ sample from Zapf et al.~\cite{Jannis}. The results are
shown in Fig.~\ref{fig:magnetization} for decreasing magnetic field. The
overall behavior is similar, but we found a saturation magnetization of
$6.82\,\mu_{\text{B}}$, in good agreement with the theoretical expectations
and Ref.~\cite{Jiang09}. 

The step-like increase of the magnetization
around $0.65\,\mu_{\text{B}}\text{T}$ was associated with a spin-flip
transition of the Eu$^{2+}$ moments in Ref. \cite{Jannis}. However,
magnetostriction, magneto-transport and unpublished neutron diffraction data
indicate that the feature is associated with the reorientation of domains,
rather than an intrinsic spin-flip of the a-twin domain and must, therefore,
be identified with $H_{1}$. Furthermore, the Eu saturation field of the b-twin
domain can be extrapolated to $H_{b}^\text{sat}\approx1.14\,$T from the slope of
the low-field region between 0.2\thinspace T and 0.5\thinspace T
(Fig.~\ref{fig:magnetization}); note that this field needs to be extrapolated,
and cannot be measured directly because at $H\gtrsim1$ T virtually all domains
are a-twins (Fig.~\ref{fig:dynamics}g). The constants can be extracted from these two fields using 
the following set of equations:%
\begin{align}
J_{\perp}  &  =\frac{M}{8}H_{b}^\text{sat}\left[ 1+2\left( \frac{H_{1}}{H_{b}%
^\text{sat}}\right) ^{2}\right] \nonumber\\
K  &  =\frac{M}{32}H_{b}^\text{sat}\left[ 1-2\left( \frac{H_{1}}{H_{b}^\text{sat}%
}\right) ^{2}\right] ,\label{eq:constants}%
\end{align}
which yields $J_{\perp}=0.093(11)$\thinspace meV and $K= 0.0049(18)$ \thinspace 
meV for this particular sample. 

The determination of the coupling constants via a different set of equations is 
discussed in the supplement. The respective results are summarized there in Tab.~(\textbf{SII}). We 
have cross-checked the results by determining the parameters from various 
samples grown with different methods and investigated with various measurements techniques like 
magnetostriction, magneto- transport and neutron diffraction data. The 
results appear to be consistent between measurements, but we found a 
noticeable sample dependence, which also seems to be related to the synthesis 
methods of the single crystals, see also Tab.~(\textbf{SII}). The averaged 
values are $J_{\perp}=0.121(24)$\thinspace meV, $K= 0.0072(22)$ \thinspace meV 
and $J_{\perp}/K\approx18(4)$. However, since each Eu atom is surrounded by 8 Fe 
atoms, $8K = 0.057$\thinspace meV is the more representative quantity to gauge the 
biquadratic coupling strength in the system.

\begin{figure}
\includegraphics[width=0.5075\textwidth]{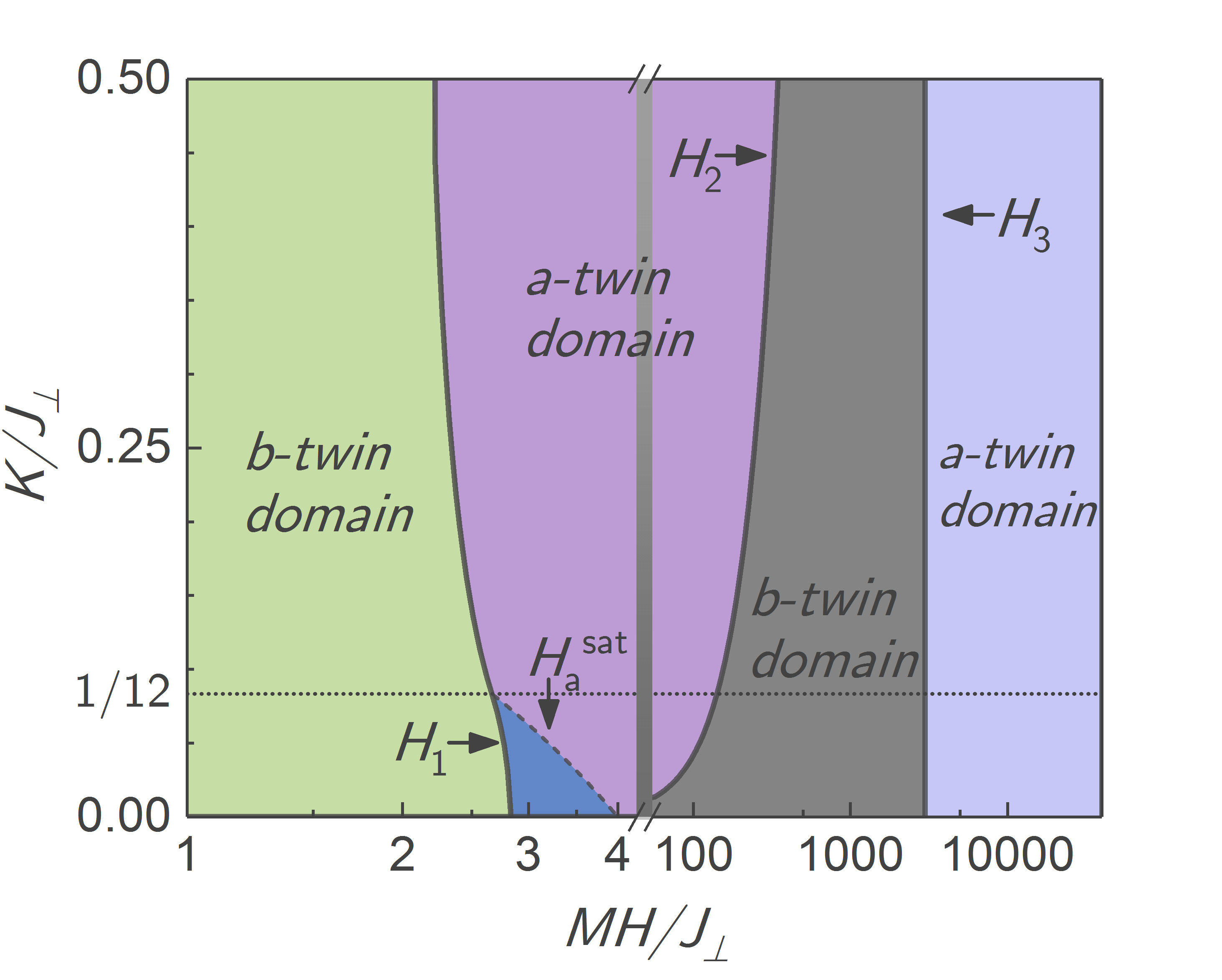}\caption{Phase
diagram $K/J_{\perp}$ vs. $M H/J_{\perp}$ in thermodynamic equilibrium. The
four phases shown correspond to: The b-twin domains (green) with
canted Eu$^{2+}$ spins, a-twin domains with saturated (purple) or unsaturated
(blue) Eu magnetization, respectively, followed by the b-twin
domains (gray) induced by the canting of Fe spins. Finally another hypothetical phase
of a-twin domains, induced by further canting of the Fe spins, is shown (light blue). 
More details, also on the significance of the dotted line, indicating $12K/J_{\perp}= 1$, 
can be found in the Supplement.}
\label{fig:phase-diag}
\end{figure}

\textbf{Energy barrier and domain dynamics} The domain dynamics are driven by
the energy difference $dE$ between the domains. In Fig.~\ref{fig:dE} we show
$dE$ on a semi-log plot as a function of the reduced variable $MH/J_{\perp}$
for a representative ratio of $K/J_{\perp}=0.05$. Positive values correspond
to the b-twin domain being the ground state, while for negative ones the
a-twin is the ground state. The phase diagram in thermodynamic equilibrium over
a large parameter space in reduced coordinates \{$K/J_{\perp}$, $MH/J_{\perp}\}$ 
is shown in Fig.~\ref{fig:phase-diag}. The remainder of the formulas used in 
constructing this phase diagram can be found in the supplement.

Until now we have assumed that the reorientation of twin domains has no energy
cost. In reality, however, there is an unknown energy barrier $\Delta$
associated with the reorientation, i.e. the energy difference
$dE=E_{\text{a-twin}}-E_{\text{b-twin}}$ between the two twin variants needs
to exceed a certain threshold before reorientation occurs. In the following,
we will assume $\Delta$ for various domain walls to be log-normal distributed,
i.e. $\log{\Delta}$ is normal\cite{Limpert01}. This is a typical distribution
e.g. of grain sizes in polycrystalline matter\cite{book}. The cumulative
distribution function $F_{X}(x)$ of a positive log-normal distributed variable
$X$ is given by $F_{X}(x)=\frac{1}{2}\erfc\left(  -\frac{\log{x}-\mu}%
{\sigma\sqrt{2}}\right)  $, with the location and the scaling parameters $\mu$
and $\sigma$. Due to the fact that $dE$ changes sign as a function of applied
field, the log-normal distribution in our case leads to the following fit function to the (a-twin) domain population:
\[
n(dE)=\left\{
\begin{array}
[c]{lr}%
n_{0}\cdot\frac{1}{2}\erfc\left(  \frac{\log(dE/\Delta)}{\sqrt{2}}\right)  , &
dE>0\\
dn\cdot\frac{1}{2}\erfc\left(  -\frac{\log(-dE/\Delta)}{\sqrt{2}}\right)
+n_{0}, & dE<0
\end{array}
,\right.
\]
with $n_{0}$ the fraction of a-twins at $H=0$ and $dn=n_{\text{sat}%
}-n_{0}$ the difference between the saturated a-twins $n_{\text{sat}}$ and
$n_{0}$. Together with Eq. \eqref{dEall} this function fully describes the
domains population.

Prior to performing the fit, the domain population needs to be extracted from
the available measurements. The domain population $n(H)$ can be determined in
a variety of ways. Arguably the most exact data can be extracted from
field-dependent neutron diffraction measurements on a free standing sample,
which will be published soon\cite{tbp}. In the following we will use
magnetostriction (MS) data by Zapf et al. \cite{Jannis}, which agree with the
preliminary data of Ref.~\cite{tbp}.

\begin{figure}
\includegraphics[width=0.5\textwidth]{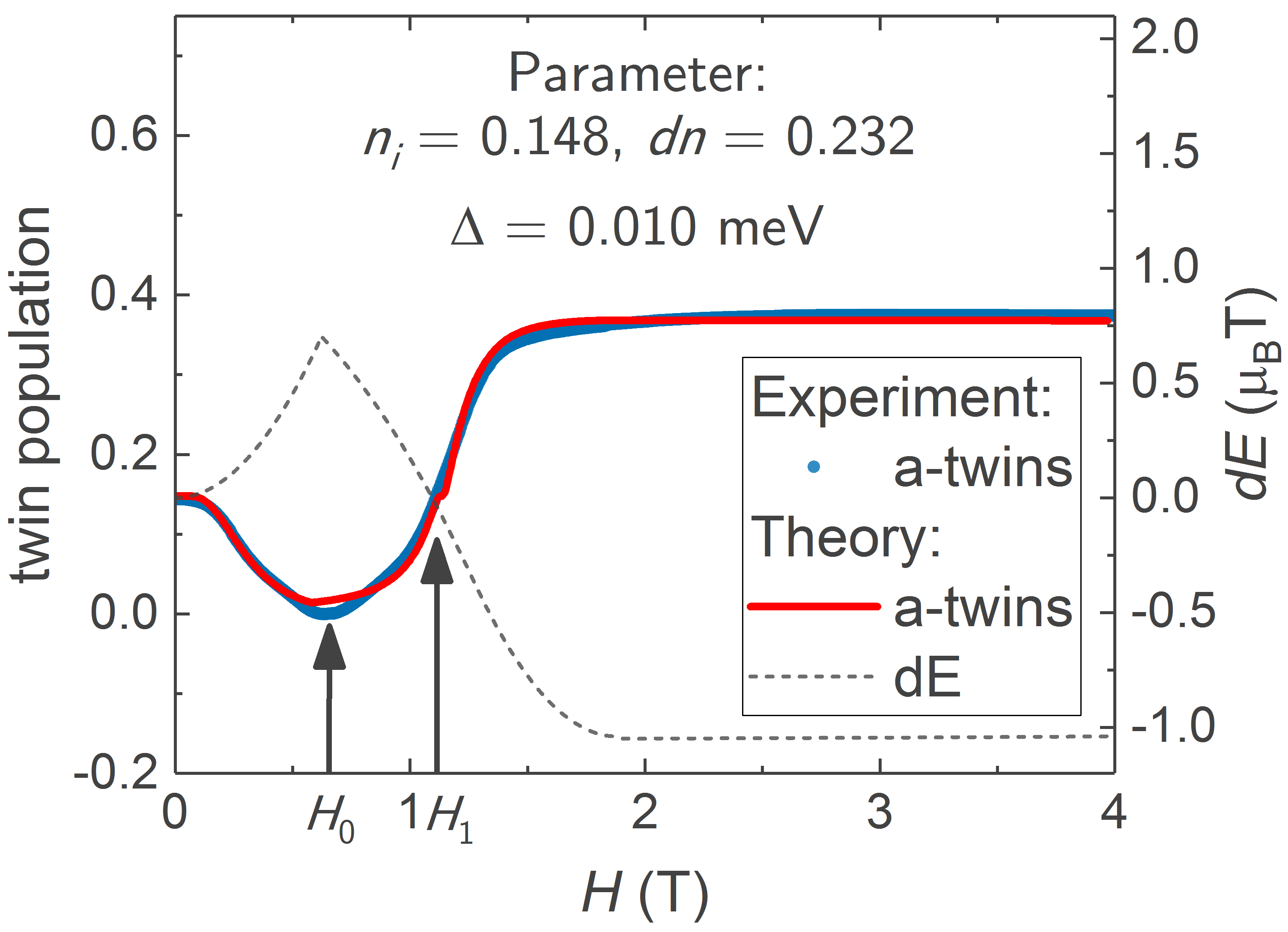}\caption{Twin
domain population derived from magnetostriction data at ($T=5K$) from Ref.~\cite{Jannis} as a function
of (increasing) magnetic field applied along the $[110]_\text{T}$ direction (blue line) as
discussed in the main text. The solid line (red) represents the theoretical
prediction using the constants' values derived in the main text and the Supplement (see Tab.\,\textbf{SII}, \#2). The dashed
line (gray) depicts the theoretically derived domain energy difference $dE$
similar to Fig.~\ref{fig:dE}(b). The a-twin domain population is reduced, due to the pressure of the dilatometer, which favors
the b-twin domain in this orientation.}%
\label{fig:MS}%
\end{figure}

The MS is defined as $\Delta L(H)/L_{0}=(L(H)-L_{0})/L_{0}$, where $L_{0}$ and
$L(H)$ are the initial and field-dependent average unit cell length,
respectively. They can be expressed by $L(H)=n(H)a+[1-n(H)]b$ and $L_{0}%
=n_{0}a+(1-n_{0})b,$ again with the (initial) domain population $n_{0}$ at
$H=0$ and orthorhombic lattice parameters $a$ and $b$. Solving for $n$ leads
to
\[
n(H)=\frac{\Delta L}{L_{0}}\left(  \frac{b}{a-b}+n_{0}\right)  ,
\]
from which follows
\begin{equation}
n_{0}=\frac{a\cdot n(H)-b[n(H)+\Delta L/L_{0}]}{(a-b)(1+\Delta L/L_{0})},
\end{equation}
assuming full detwinning at the observed minimum around $H_{0}$,
with $b||H$, i.e. $n(H_{0})=0$. This assumption is justified, as
preliminary neutron data\cite{tbp} indicate a domain distribution with
$n(H_{0})=0.06$. Furthermore, the significant pressure of the
dilatometer in field direction, 1.35\thinspace MPa\cite{Jannis} aids the $n=0$
alignment at small magnetic fields (but hinders it at larger fields, when
$n\rightarrow1$). The extracted domain population is shown in
Fig.~\ref{fig:MS}. The solid red line represents the fit to the data (blue
symbols). The energy barrier for this particular sample and measurement
technique was determined to $\Delta=0.01$\thinspace meV. Among the other investigated samples
$\Delta$ ranges roughly between $10^{-3}$ meV and $10^{-2}$ meV, in agreement
with the estimates presented in the introduction.

\textbf{Consistency check}

Using the averaged constants from the previous paragraph we find $H_{1}=0.85
$\thinspace T and ${H}_{2}=35$\thinspace T, in very good agreement with
experiment. Utilizing the knowledge about the energy barrier, we can also calculate
the initial detwinning field from the condition $dE=\Delta$, through
\[
H_{0}=\frac{2}{M}\sqrt{2\Delta(J_{\perp}+4K)}.
\]
For the determined $\Delta$-range between 0.001\thinspace meV and 0.01\thinspace meV
this yields 0.09\thinspace T to 0.28\thinspace T which is also in excellent
agreement with the experimental evidence.

Going even further, the presented theory allows us to calculate our new
magnetization data, by weighting the Eu magnetization of the twin sublattices
with the domain population $n(H)$ (dashed line in Fig.~\ref{fig:magnetization}), that we calculated using
$\Delta=0.004$\thinspace meV, $n_{i}=0.22$ and $dn=0.71$. The total
magnetization is given by
\begin{equation}
M(H)=n(H)\cdot M_{a}+(1-n(H))\cdot M_{b},
\end{equation}
where $M_{a}$ and $M_{b}$ are the magnetization of the sublattices given by
$M\cos{\varphi}$, i.e.
\begin{align*}
M_{a}&=\left\{
\begin{array}
[c]{l}%
\left.
\begin{array}
[c]{ll}%
0, & \{0,H_{a}^{\text{flop}}\}\\
\frac{M^{2}H}{4(J_{\perp}-4K)}, & \{H_{a}^{\text{flop}},H_{a}^{\text{sat}}\}\\
M, & \{H_{a}^{\text{sat}},\infty\},\\
&
\end{array}
\right. \\
\end{array}
\right. \\
M_{b}&=\left\{
\begin{array}
[c]{l}%
\left.
\begin{array}
[c]{ll}%
\frac{M^{2}H}{4(J_{\perp}+4K)}, & \{0,H_{b}^{\text{sat}}\}\\
M, & \{H_{b}^{\text{sat}},\infty\}.\\
&
\end{array}
\right. \\
\end{array}
\right.
\end{align*}
The result is depicted by the solid red line in
Fig.~\ref{fig:magnetization}.

\textbf{Summary} We present a microscopic, physically meaningful and
quantitative description of the observed magnetic detwinning effect in FeBS,
with all its complexity. In particular, the following mysteries have been
resolved: (i) strong detwinning in minuscule fields despite absence of
spin-orbit coupling effects in Eu$^{2+}$ ions; (ii) coupling of Eu spin
orientation to the Fe sublattice, despite any bilinear interaction canceling
out by symmetry, and (iii) double reversal of the preferential domain
orientation with the increase of the external field. We show that all these
issues find a natural explanation if a \textit{biquadratic }Eu-Fe coupling is
included in the model. We also show that such a term does actually appear in
first-principles calculation, with an amplitude even stronger than needed to
explain the experimental data. Furthermore, we were able to describe
quantitatively not only the thermodynamic phase diagram, but even the dynamics
of detwinning (as deduced from our new magnetization data), assuming that the
detwinning energy barriers are distributed according the log-normal law (quite
typical in crystal morphology). 

Although we focus on the Eu-based 122-system, our findings should be of great
interest for other large-spin rare earths as well, such as the Gd-based 1111-compound,
which, like EuFe$_2$As$_2$, also features large $S$=7/2 spin-only moments, or the recently
discovered Eu-based 1144-systems. Furthermore, the theory also captures the physics of the 
high-field ($H\agt 15$\, T) detwinning, which occurs even in non Eu-based iron pnictides.
The microscopic understanding of
the phenomenon of magnetic detwinning in ultra-low magnetic fields, as deduced above,
opens new avenues for the experimental investigation of spin-driven nematicity in Fe-based superconductors.

\textbf{Acknowledgments}

We are grateful to Shibabrata Nandi and Yinguo Xiao for giving us access to 
their neutron diffraction data, to Makariy Tanatar for sharing with us his 
data on mechanical detwinning and to Ian Fisher for discussing with us the 
concept of this work. We also like to convey our gratitude to Shuai Jiang, 
Christian Stingl and Jeevan H.S. for permitting us access to their samples 
and magnetostriction data and Sina Zapf and Martin Dressel for collaboration 
in the early stage of this project (Ref.~\cite{Jannis}), and to James Glasbrenner
for verifying our biquadratic calculations using an all-electron method. J.M. thanks Patrick 
Seiler for support and discussions. J.M. and P.G. are supported by DFG 
through SPP 1458. I.I.M. is supported by ONR through the NRL basic research 
program.


\begin{thebibliography}{26}
\bibitem [{not()}]{note}%
  \BibitemOpen
  \href@noop {} {\enquote {\bibinfo {title} {It is well established that a
  microscopically justifiable model includes heisenberg terms up to the third
  neighbors and a sizable n.n biquadratic coupling \cite{Kirill}, but for our
  purpose of calculating the spin canting in an external magnetic field and the
  corresponding energy gain, all this complexity can be absorbed in one
  parameter, of which we only need to know the order of magnitude.}}\
  }\BibitemShut {NoStop}%
\end{thebibliography}

%

\end{document}


\title{Microscopic theory of magnetic detwinning in \\ iron-based superconductors with large-spin rare earths\\-- Supplementary Information --}
\author{Jannis Maiwald}
\affiliation{Experimentalphysik VI, Universit\"at Augsburg, Universit\"atstra{\ss }e 1,
86135 Augsburg, Germany}
\author{I. I. Mazin}
\affiliation{Code 6393, Naval Research Laboratory, Washington, DC 20375, USA}
\author{Philipp Gegenwart}
\affiliation{Experimentalphysik VI, Universit\"at Augsburg, Universit\"atstra{\ss }e 1,
86135 Augsburg, Germany}
\date{\today}
\maketitle



\section{Methods}
\subsection{Computations}
In order to estimate computationally the key parameters of the theory (keeping
in mind that the final values would be extracted from the experiment), we have
performed calculations using the plane wave code VASP\cite{Kresse} with the
setup as described in Ref. \cite{setup}, and using the Generalized Gradient
Approximation (GGA) to the Density Functional Theory (DFT). Our first goal was
to verify the conjecture based on the general considerations presented in the main text, that 
the single-site anisotropy (SSA) for Eu can be neglected. This is not a trivial task:
 rotating both Eu and Fe spins together incurs a much larger Fe SSA, while rotating only
Eu  entangles Eu SSA with the Fe-Fe biquadratic coupling (and may be unreliable for
different reasons, as discussed below). We can assess the SSA anisotropy of Eu indirectly though,
bypassing the problem of Eu-Fe coupling entirely. 
We can replace Fe$^{2+}$ with the isovalent, but nonmagnetic Zn$^{2+}$. The EuAs trilayers 
then remain intact, and if there is any coupling of Eu magnetic energy with the orthorhombic
distortion of the Eu sublattice, it should manifest itself in EuZn$_2$As$_2$ as well. Everything else is
interaction between Eu spins and Fe spins, which can be described as a combination of a Heisenberg (bilinear)
and a biquadratic coupling. We have performed this calculation using both the VASP code and the all-electron
Linear Augmented Plane Waves WIEN2k code\cite{WIEN}, and obtain the SSA of 0$\pm$ 0.007 meV, 
with the error window an order of magnitude smaller than required to explain the experiment.
We can get an independent estimate of the Eu SSA by going back to our compound and fixing the Fe spins
along the $c$ axis, while rotating Eu spins in the $xy$ plane. This way, the biquadratic coupling between Eu and
Fe remains always zero. In these calculations we found the total energy to be independent of the Eu orientation
within a similar error bar, thus confirming that the Eu closed shell does not allow for any SSA.

We also note that possible anisotropic exchange interactions of the form 
$W(M_{ij}^{x}M_{i+1,j}^{x}-M_{ij}^{y}M_{i,j+1}^{y})$ contribute to the calculated energy differences in both
 these procedures at the same footing as SSA (since the Eu spins remain collinear), so the null
 results of the calculations also excludes these type of effects. Finally, the Dzyaloshinskii-Moria
 interaction in this system is also ruled out, because the midpoint between two Eu-sites is an inversion center. 
These two interaction exhaust relevent anisotropic exchange term allowed by the square-planar symmetry.

\begin{figure}
\includegraphics[width=0.5\textwidth]{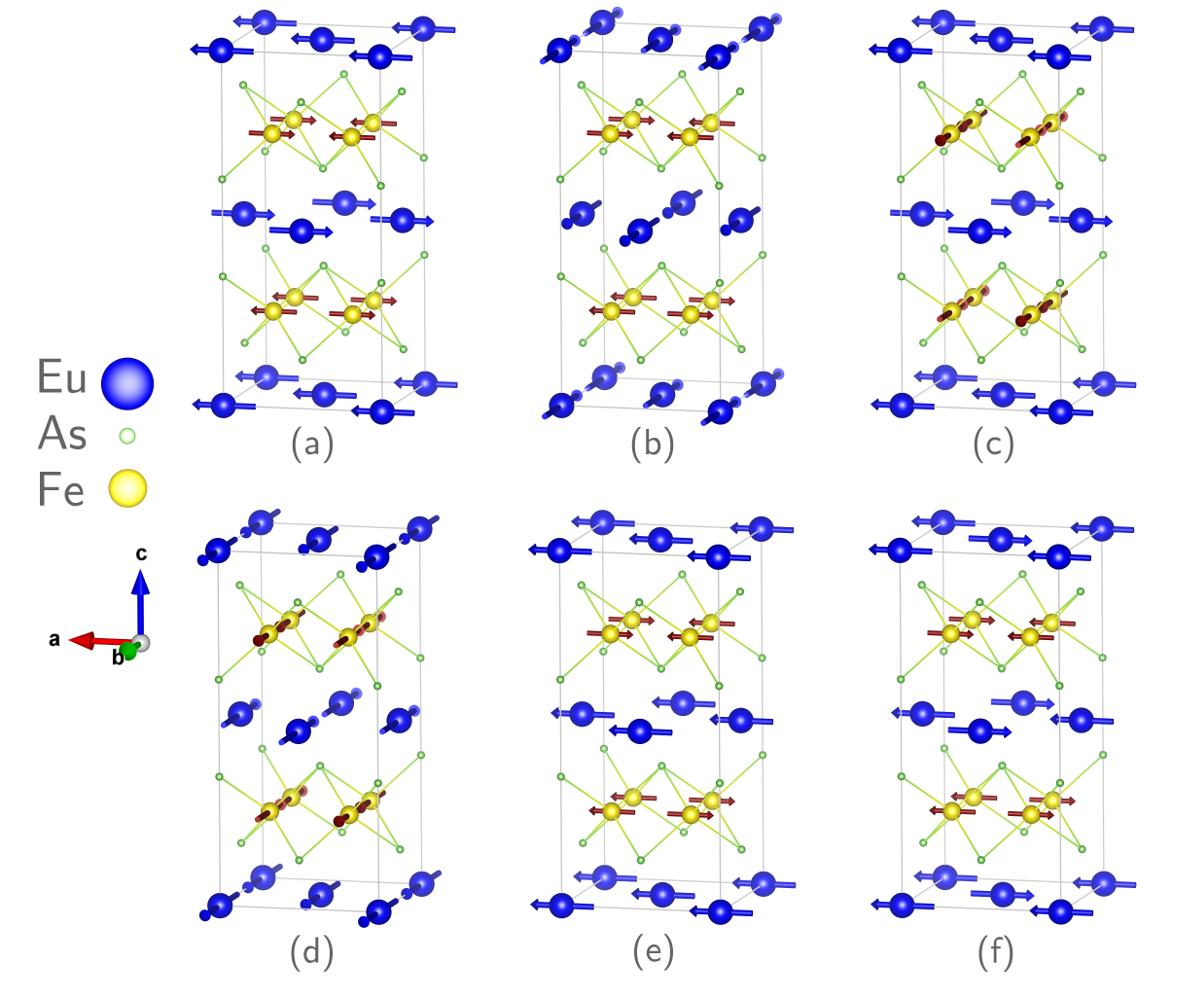}\caption{
Patterns used for estimating the model parameters from DFT calculations: noncollinear pattern pairs (a,b) and (c,d)
were used for estimating $K$, collinear pairs (a,e) for $J_\perp$, and (a,d) for $J_{||}$.}%
\label{fig:patterns}%
\end{figure}

We have also tried to calculate the biquadratic coupling $K$ from first principles. To this end,
we have performed two sets of calculations: one with Fe
spins along the easy axis $a,$ and Eu spins either parallel or perpendicular
to the former, and then taking the energy difference, or with Fe spins along 
$b,$ and Eu spins either parallel or perpendicular to the former,
again taking the energy difference. Not unexpectedly, the result depends on
the Hubbard $U,$ but is consistent between the two sets for the same $U$. In
particular, for $U-J=7$ eV, we found $K\approx0.4$ meV. We have also performed calculations gradually 
rotating the Eu moments with respect to the Fe ones,
using the constrained-magnetic-moment formalism, incorporated in VASP. We have obtained
the total energy roughly proportional to $\cos ^2 \phi$, where $\phi$ is the angle
between the Fe and Eu spins, with the coefficient consistent with the above estimate.
Compared to the values extracted from the
experiment, these numbers are too large, partially because the GGA
underestimates the wave function localization, and the biquadratic term is small and very
 sensitive to localization. However, the discrepancy is still too large.
It should be noted that while collinear LDA+U+SO calculations have a solid history
by now, and have been tested for many materials, including those containing rare earths,
the same cannot be said about noncollinear calculations. Proper account of the 
nonspherical part of the double-counting term in LDA+U is very challenging\cite{Anisimov97} . Because of that, VASP developers
recommend using a spherically averaged potential, replacing $U$ with $U-J$ and setting $J$ to zero.
This is what we did, and, in fact, we are not aware of any publicly available DFT
package that would allow for noncollinear magnetic calculation and properly
included nonspherical LDA+U. With this in mind, we have tested the stability
of the above-mentioned coefficient in front of $\cos ^2 \phi$ with respect to $U$.
We found that it depends unphysically strongly on $U$.
This means that while calculations can be taken as an indication that the experimental, or even larger
biquadratic coupling is consistent with the band structure calculations, its amplitude cannot
be reliably estimated computationally.

Estimating $J_{\perp}$ does not require noncollinear calculations and is more reliable. 
we compared the total energies of ferromagnetic Eu planes, stacked
ferromagnetically or antiferromagnetically along c. Finally, for estimating
$J_{||}$ we compared ferromagnetic and checkerboard-antiferromagnetic in-plane
arrangements of Eu ions, always allowing stripe antiferromagnetism for Eu. All
these patterns are illustrated in Fig.~\ref{fig:patterns}.

\subsubsection{The role of bilinear coupling between Fe and Eu}
The fact that the Heisenberg coupling induces zero exchange field at the Eu sites (which
project onto the centers of Fe$_4$ plaquettes) from the stripe-ordered
Fe planes is quite obvious. One can show that not only Heisenberg exchange, 
but any symmetry-allowed bilinear spin-spin coupling generates at the Eu sites,
at best, an exchange field perpendicular to the Fe planes. In fact,
since the symmetry positions of the As atoms with respect to the Fe planes are
exactly the same as the Eu atoms, the following derivation exactly
follows the well known theory of the As hyperfine field
in BaFe$_2$As$_2$ in the NMR community (see, for instance Ref.~\cite{Ok17}) .

Let us assume an arbitrary bilinear coupling between Fe and Eu, of
the form, for a single bond%
\begin{equation}
H=e_{\alpha}T_{\alpha\beta}f_{\beta}, \nonumber
\end{equation}
with $\mathbf{e}$ and $\mathbf{f}$ the Eu
and Fe spins, respectively and summation over repeated Cartesian indices. The structure of $T$ depends on the direction
of the bond. The Heisenberg exchange corresponds to $T=c \cdot\delta
_{\alpha\beta},$ dipole interaction to $c\cdot(3d_{\alpha}d_{\beta}%
-d^{2}\delta_{\alpha\beta}),$ the DM interaction to $c \cdot\varepsilon
_{\alpha\beta\gamma}d_{\gamma}$ (equivalent to $\mathbf{d\cdot e\times f)}$, where $c$ is a constant and $\mathbf{d}$ the Eu-Fe bond vector.
In general there are symmetry constraints on allowed elements of the $T$
matrix, but our derivation below will not be using them explicitly.

We consider a plaquette of four Fe atoms at $\{\pm1,\pm1,0\}$ (site labels starting from \{1,1,0\} counting counterclockwise), with their
spins along $x$ as $\{1,0,0\},$ $\{-1,0,0\},$ $\{-1,0,0\},$ $\{1,0,0\},$ and
Eu at $\{0,0,h\}.$ The exchange field generated by the Fe$_{1}$ at the Eu site
is
\begin{equation}
h_{\alpha}^{(1)}=T_{\alpha\beta}^{(1)}f_{\beta}^{(1)}=T_{\alpha x}^{(1)} \nonumber%
\end{equation}
In the following we show how the exchange fields of the other sites can be expressed in terms of the Fe$_1$ site with spin direction either along \{1,0,0\} or along \{0,1,0\}. The Fe$_{2}$ site e.g. is obtained by rotating a \{0,1,0\} spin at the Fe$_{1}$ site (corresponding to $h_{\alpha}=T_{\alpha y}^{(1)}$) around the
$z$ axis by $\pi/2.$ This rotation is given by the matrix%
\begin{equation}
R^{(2)}=\left(
\begin{array}
[c]{ccc}%
0 & -1 & 0\\
1 & 0 & 0\\
0 & 0 & 1
\end{array} \nonumber
\right).
\end{equation}
After the rotation we get a spin along $\{-1,0,0\}=\mathbf{f}^{(2)}$ at site 2, generating the field $\mathbf{h}^{(2)}=R^{(2)}\cdot\mathbf{h}=R_{\alpha\beta}^{(2)}T_{\beta
y}^{(1)}$. By symmetry, this must be the same as the field generated by
$\mathbf{f}^{(2)}$ through the matrix $T^{(2)},$ namely $T_{\alpha x}^{(2)}.$
Thus, $T_{\alpha x}^{(2)}=R_{\alpha\beta}^{(2)}T_{\beta y}^{(1)}.$ Using
similar arguments, we find that
\begin{align}
h_{\alpha}^{(1)}  & =\delta_{\alpha\beta}T_{\beta x}^{(1)}, \hspace{1cm}
h_{\alpha}^{(2)}   =R_{\alpha\beta}^{(2)}T_{\beta y}^{(1)}\nonumber \\
h_{\alpha}^{(3)}  & =R_{\alpha\beta}^{(3)}T_{\beta x}^{(1)}, \hspace{0.87cm}
h_{\alpha}^{(4)}   =R_{\alpha\beta}^{(4)}T_{\beta y}^{(1)},\nonumber %
\end{align}
where
\begin{equation}
R^{(3)}=\left(
\begin{array}
[c]{ccc}%
-1 & 0 & 0\\
0 & -1 & 0\\
0 & 0 & 1
\end{array}
\right)  ,\;R^{(4)}=\left(
\begin{array}
[c]{ccc}%
0 & 1 & 0\\
-1 & 0 & 0\\
0 & 0 & 1
\end{array}
\right)  . \nonumber
\end{equation}
Adding up the four fields, leads to:%
\begin{align}
h_{\alpha}^\text{tot}  & =\left(\delta_{\alpha\beta}+R_{\alpha\beta}^{(3)}\right)T_{\beta
x}^{(1)}+\left(R_{\alpha\beta}^{(2)}+R_{\alpha\beta}^{(4)}\right)T_{\beta y}^{(1)}\nonumber \\
& =\left(
\begin{array}
[c]{ccc}%
0 & 0 & 0\\
0 & 0 & 0\\
0 & 0 & 2
\end{array}
\right)  \left(T_{\beta x}^{(1)}+T_{\beta y}^{(1)}\right)\nonumber \\
& =\{0,0,2(T_{zx}^{(1)}+T_{zy}^{(1)})\}, \nonumber
\end{align}
which does not have any component in the $xy$ plane. If the interaction in
question is dipole, this is similar to the well-known derivation of the NMR
hyperfine field on As. In case of the DM interaction, the last term also cancels and the entire
field is zero.

\subsection{Experiment}

Single crystals of EuFe$_2$As$_2$ were grown using three different techniques, namely,
 Fe-As self-flux\cite{Jiang09}, Sn-flux and the Bridgman method\cite{Jeevan11}. The respective sample preparation
 methods were similar to those in the corresponding references. 
Samples were characterized using energy-dispersive x-ray analysis and x-ray diffraction to confirm composition and
 structure. Afterwards, they were oriented using a Laue camera and subsequently cut along the tetragonal [110]$_\text{T}$-direction with an electric spark erosion technique. Typical dimensions after cutting were approximately 2$\times$1$\times$0.1\,mm$^3$.

Magnetization measurements were performed in a standard Magnetic Property Measurements System (MPMS), while DC-transport data was recorded using a standard Physical Property Measurement System (PPMS) in the Montgomery geometry.

\section{Model Details}

In the main text, we have determined the {\it equilibrium} phase diagram. Here, we shall derive the energies of each 
type of domain for all fields, so as to be able to deduce the energy cost/gain of any reorientation. Eq.~12 
in the main text is based on this derivation. We will also present, for completeness, the energetics at ultra-high
 fields, not accessible in current experiments.

We start from Eq.~4 of the main text, and will first calculate the energy of the b-twin ($\mathbf{H}\parallel b$), and then of the a-twin, in all fields. We will consider separately two different
regimes: first, small fields compared to $H_2$ defined in the main text, and, second, large fields 
comparable with, or larger than $H_2$. In the former regime we can safely neglect the Fe spins' canting ($\tilde{\varphi}_2
=\tilde{\varphi}_1 =\pi/2$, $\tilde{\alpha}_2=\tilde{\alpha}_1=0$, while in the latter we can use the fact that the
Eu magnetization is already saturated and $\varphi_1=\varphi_2=0$. 

\subsection{b-Domain:  $\mathbf{H}\parallel b$.}
\subsubsection{Small Fields}
At \textit{small fields} we have, $\tilde{\varphi}_2=\tilde{\varphi}_1
=\pi/2$, $\tilde{\alpha}_2=\tilde{\alpha}_1=0$ and $\varphi=\varphi_1=\varphi_2$, which, when
 inserted into Eq.~(4) of the  main text leads to
\begin{eqnarray}
E_{b}&=&-MH\cos\varphi+J_{\perp}\cos2\varphi-8K\sin^{2}\varphi
-2\tilde{J}-2\tilde{D} \nonumber \\
&=&-MHp+(2J_{\perp}+8K)p^{2}+E_{0}, 
\label{eq:atwin}
\end{eqnarray}
with $p=\cos\varphi$ and $E_{0}=-2\tilde{J}-2\tilde{D}-J_{\perp}-8K$.
Minimizing Eq.~\eqref{eq:atwin} with respect to $p$, leads for $MH<4(J_{\perp}+4K)$ to the equilibrium tilting 
angle and energy given in Eq.~(6-7) of the main text. For sufficiently small magnetic
fields this solution is always below $E_{0}$ and Eu$^{2+}$ moments can screen the field continuously. The tilting angle $\varphi$ changes gradually according to 
\begin{equation}
\varphi=\cos^{-1}\left[\frac{MH}{4J_{\perp}+16K}\right].
\label{eq:phi_b}
\end{equation}
At fields
larger than $H_{\text{b}}^{\text{sat}}=(4J_{\perp}+16K)/M$ (derived from $p=1$) the moments are fully
aligned with the external magnetic field and the energy changes to
\begin{equation}
E_{\text{b}}^{\text{sat}}=-MH+2J_{\perp}+E_{0}+8K.
\label{eq:E_B_min2}%
\end{equation}
\subsubsection{Large Fields}
At considerably higher fields the Fe moments will start to
deflect ($\tilde{\alpha}=\pi/2-\tilde{\varphi}$) from the ground state orientation, while the Eu$^{2+}$ moments are already saturated. In this case Eq.~(4) of the main text yields 
\begin{align*}
\tilde{E}_{b}   =&-2\tilde{M}H\cos\tilde{\varphi}+2\tilde{J}%
\cos2\tilde{\varphi}-2\tilde{D}\sin^{2}\tilde{\varphi}\\
&  -8K\cos^{2}\tilde{\varphi}-MH+J_{\perp}\\
=&-2\tilde{M}H\tilde{p}+2(2\tilde{J}+\tilde{D}-4K)\tilde{p}^{2}\\
&  -MH+E_{0}+2J_{\perp}+8K
\end{align*}
where we have now introduced $\tilde{p}=\cos\tilde{\varphi}$ and $E_{0}$. The equilibrium energy is now determined by minimizing with respect to $\tilde{p}$, which then reads
\begin{equation}
\tilde{E}^\text{min}_{\text{b}}=-MH+2J_{\perp}+8K-\frac{(\tilde{M}H)^{2}}{2(2\tilde{J}\nonumber
+\tilde{D}-4K)}+E_{0},\label{eq:E_B_min2+}%
\end{equation}
i.e. Eq.~\eqref{eq:E_B_min2} with an additional term stemming from the canting
of the Fe moments away from the easy axis. 

\begin{figure}
\includegraphics[width=0.5\textwidth]{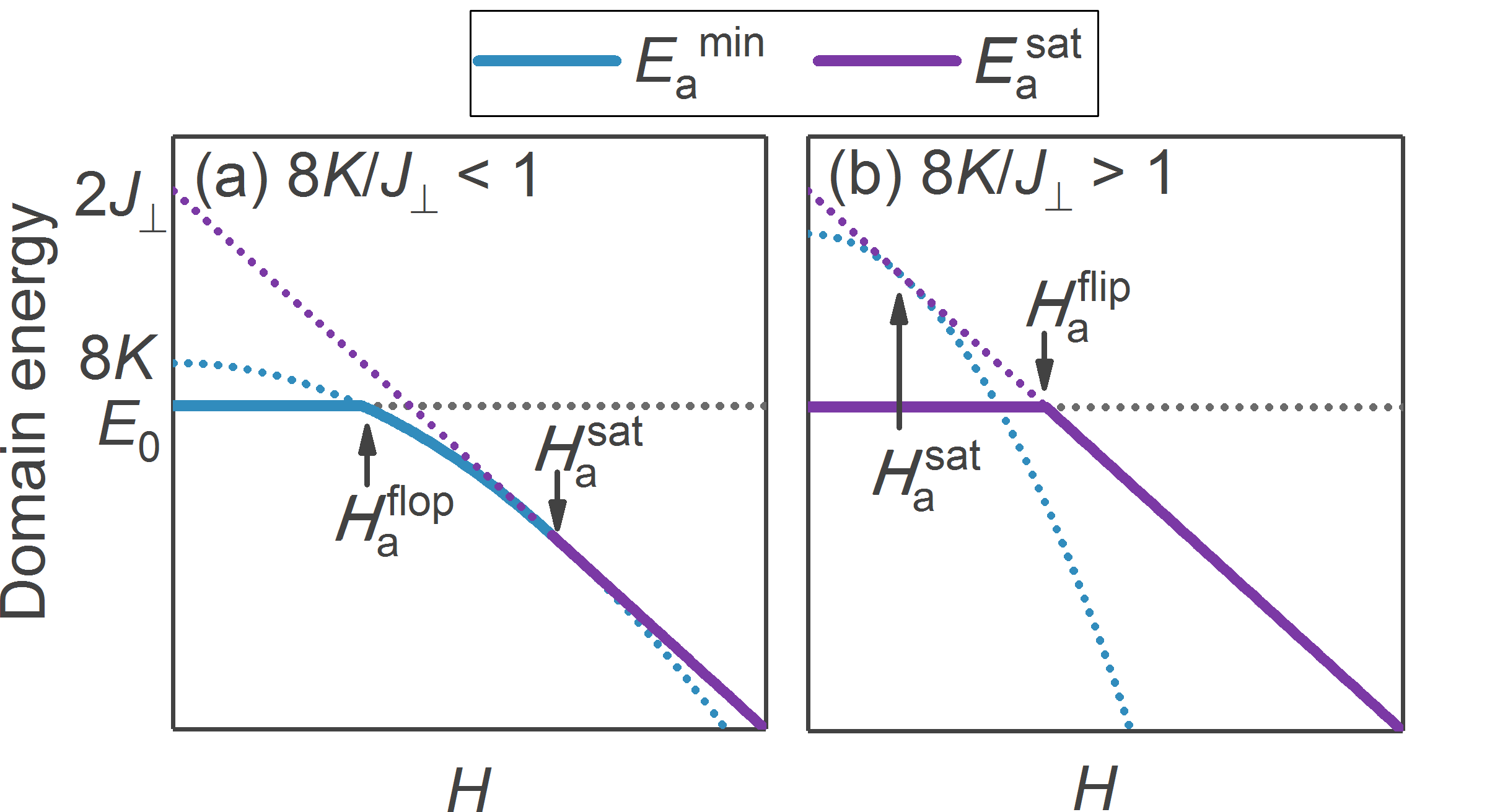}\caption{Calculated a-twin
domain energy of EuFe$_{2}$As$_{2}$ as a function of the applied magnetic
field for (a) the spin-flop and (b) the spin-flip case. }%
\label{fig:aenergy}%
\end{figure}

\subsection{a-Domain: $\mathbf{H}\parallel a$}
\subsubsection{Small Fields}
At small fields we have the following conditions: $\tilde{\alpha}_1=\tilde{\varphi}_1 = 0$,
$\tilde{\alpha}_2=\tilde{\varphi}_2 = \pi$ and $\varphi=\varphi_1=\varphi_2$, leading to 
\begin{align}
E_{a}  
& =-MH\cos\varphi+J_{\perp}\cos2\varphi-8K\cos^{2}%
\varphi-2\tilde{J}-2\tilde{D}\nonumber\\
&  =-MHp+(2J_{\perp}-8K)p^{2}+E_{0}+8K,\nonumber
\label{eq:btwin}
\end{align}
again with $p=\cos\varphi$ and $E_{0}$, which after minimizing w.r.t. $p$ results for
$J_{\perp}>4K$ and $MH<4J_{\perp}-16K$ in Eqs.~(9-10) of the main text. We note the sign change in the denominator and the additional $8K$-term. Due to this term this state will only be populated once $H$ is large enough. At smaller fields the Eu$^{2+}$ moments will remain in their ground state configuration. The field at which the moments spin-flop $H_{\text{a}}^{\text{flop}}=8/M\sqrt{K(J_{\perp}-4K)}$ is determined from the
condition $E^\text{min}_{\text{a}}=E_{0}$. In the spin-flop phase the Eu$^{2+}$ moments gradually rotate towards saturation according to 
\begin{equation}
\varphi=\cos^{-1}\left[\frac{MH}{4J_{\perp}-16K}\right].
\label{eq:phi_a}
\end{equation} 
Saturation is reached at $H_{\text{a}}^{\text{sat}}=(4J_{\perp
}-16K)/M$ at which point the energy in Eq.~(10) of the main text changes to
\begin{equation}
E_{\text{a}}^{\text{sat}}=-MH+2J_{\perp}+E_{0}.\nonumber
\label{eq:E_A_min2}%
\end{equation}

However, if the biquadratic coupling $K$ is strong compared to the
interplanar coupling $J_\perp$ between Eu$^{2+}$ moments, they directly \textit{flip} from the ground state into the fully saturated state, skipping Eq.~(10) entirely. Such a spin-flip occurs only if $E^\text{min}_{a}(H_{a}^\text{sat})\geq
E^\text{min}_{a}(H_{a}^\text{flop})=E_{0}$, that is to say when $J_{\perp}\leq8K$. The corresponding field is then given by $H_{a}^{\text{flip}}=2J_{\perp}/M$. The domain energy of the a-twin for the spin-flop and spin-flip case is shown in Fig.~\ref{fig:aenergy}.

\begin{figure}
\includegraphics[width=0.5\textwidth]{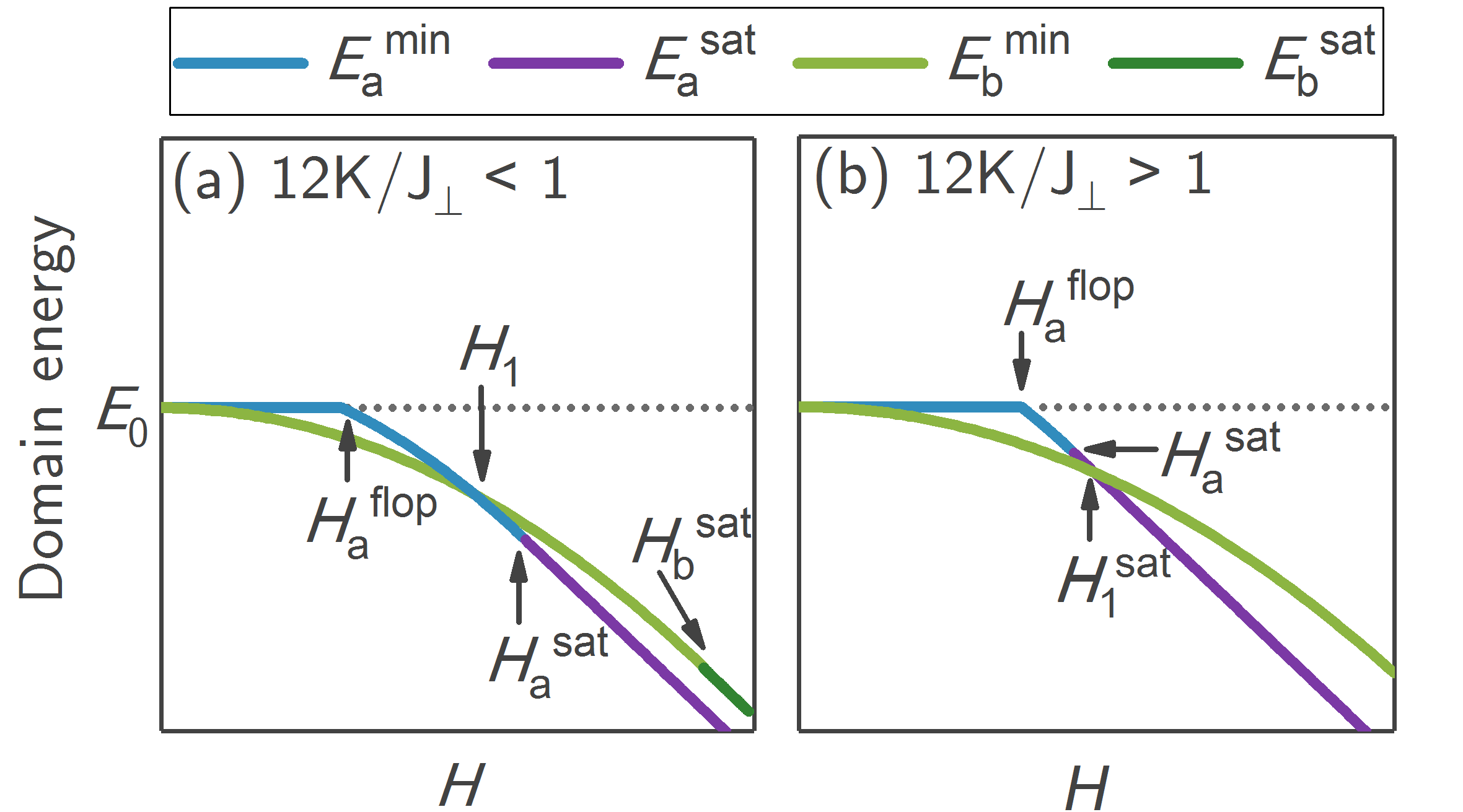}\caption{Calculated a-twin and
b-twin domain energies of EuFe$_{2}$As$_{2}$ as a function of the applied
magnetic field in two regimes (a) $12K/J_{\perp}<1$ and (b) $12K/J_{\perp}>1$.}%
\label{fig:E_ab2}%
\end{figure}

\subsubsection{Large field}

If the field is large enough the Fe moments can become subject to a spin-flop (but not a spin-flip, because $\tilde{J} \gg \tilde{D}$). As estimated in the main text this should happen around 130\,T. We can easily incorporate this effect for the sake of completeness, by setting $\varphi=0$ and $\tilde{\alpha} = \tilde{\varphi}$, which leads to

\begin{align*}
\tilde{E}_a =	&-2\tilde{M}H\cos\tilde{\varphi}+2\tilde{J}\cos 2\tilde{\varphi} -2\tilde{D}\cos^2\tilde{\varphi} \\
							&- 8K\cos^2\tilde{\varphi}-MH+J_\perp 
\end{align*}
and after replacing $\tilde{p}=\cos\tilde{\varphi}$ and $E_{0}$ to

\begin{align*}
	\tilde{E}_a	= &-2\tilde{M}H \tilde{p} + 2(2\tilde{J}-\tilde{D}-4K)\tilde{p}^2\\
							&-MH+E_0+2J_\perp+2\tilde{D}+8K,
\end{align*}
Subsequent minimization w.r.t. $\tilde{p}$ results in
\[
\tilde{p}^\text{\,min}_{\text{\,flop}}=\frac{\tilde{M}H}{2(2\tilde{J}-\tilde{D}-4K)},
\]
and
\[
\tilde{E}_a^\text{min} = -MH+2J_\perp -\frac{(\tilde{M}H)^2}{2(2\tilde{J}-\tilde{D}-4K)}+E_0 +2\tilde{D}+8K,
\]
where the Fe spin-flop field is given from the condition $\tilde{E}_a^\text{min}-E_a^\text{sat} =0$, i.e. $\frac{(\tilde{M}H)^2}{2(2\tilde{J}-\tilde{D}-4K)}=2\tilde{D}+8K$:
\[
\tilde{H}_a^\text{flop} = \frac{2}{\tilde{M}}\sqrt{(\tilde{D}+4K)(2\tilde{J}-\tilde{D}-4K)}
\]
In the saturated case it follows from $\tilde{p}=1$ that
\[
\tilde{E}_a^\text{sat} = -MH+2J_\perp -2\tilde{M}H+4\tilde{J}+E_0,
\]
which becomes valid at the saturation field 
\[
\tilde{H}_a^{sat}=\frac{2}{\tilde{M}}(2\tilde{J}-\tilde{D}-4K).
\]

\subsection{Eu spin angles at $H_1$}

In the main text we implied, for simplicity, that $\varphi(H_1) = \pi/4$ when the system changes from
b-twins to a-twins at $H_1$. However, this holds only in the limit of $K \rightarrow 0$. At finite $K$ the Eu
spin angle changes discontinuously from a value $\varphi > \pi/4$ to a value $\varphi < \pi/4$, as
can be seen when inserting the expression for $H_1$, Eq.~(11) in the main text, into Eqs.~\eqref{eq:phi_b}
and \eqref{eq:phi_a}. This corresponds to a discontinuous decrease of $\varphi$ from roughly 55$^\text{o}$
to 25$^\text{o}$ for our averaged findings of $J_\perp$ and $K$. The respective jump in
magnetization $M(H_1)$ can easily be calculated using Eq.~(15) of the main text and is in agreement with the experiment.

\begin{table*}
\caption{The energy difference $dE$ in different regimes, assuming $12K/J_\perp <1$.
In case of $12K/J_\perp >1$ and $8K/J_\perp<1$, $H_1$ changes to $H_1^\text{sat}$, which lies in the
 interval $\{H_a^\text{sat}, H_b^\text{sat}\}$ (Fig.~\ref{fig:E_ab2}), while for $8K/J_\perp>1$ a spin-flip at $H_a^\text{flip}$ is realized instead of the spin-flop at $H_a^\text{flop}$.}%
\label{tab:summary}
\begin{tabular}
[c]{r|c|r|l|c}%
\multicolumn{2}{c|}{$dE$} & \multicolumn{1}{c|}{Stability range} &
\multicolumn{1}{c|}{Critical field} & \multicolumn{1}{c}{Domain type}\\ \toprule
$E_{0}-E_{b}^\text{min}$ & $=\frac{M^{2}H^{2}}{8(J_{\perp}+4K)}$ & $0\leq H\leq
H_{a}^\text{flop}$ & $H_{a}^\text{flop}=8/M\sqrt{K(J_{\perp}-4K)}$ & b-twin\\\cline{1-3}%
\multirow{2}{*}{$E_a^\text{min}-E_b^\text{min}$} &
\multirow{2}{*}{$=K\left(8-\frac{M^2H^2}{J_\perp^2-16K^2}\right)$} &
$H_{a}^\text{flop}\leq H\leq H_{1}$ & $H_1=4/M\sqrt{(J_{\perp}^{2}-16K^{2})/2}$ &
($dE >0$)\\\cline{4-5}
&  & $H_{1}\leq H\leq H_{a}^\text{sat}$ & $ H_{a}^\text{sat}=4/M(J_{\perp}-4K)$ &
a-twin\\\cline{1-3}%
$E_{a}^\text{sat}-E_{b}^\text{min}$ & $=-M H+2J_{\perp}+\frac{M^{2}H^{2}}{8(J_{\perp}+4K)}$ &
$H_{a}^\text{sat}\leq H\leq H_{b}^\text{sat}$ & $H_{b}^\text{sat}=4/M(J_{\perp}+4K)$ & ($dE
<0$)\\\cline{1-3}%
\multirow{2}{*}{$E_a^\text{sat}-\tilde{E}_b^\text{min}$} &
\multirow{2}{*}{$=-8K+\frac{\tilde{M}^2H^2}{2(2\tilde{J}+\tilde{D}-4K)}$} & $H_{b}^\text{sat} \leq
H \leq H_{2}$ & $H_2=4/\tilde{M}\sqrt{K(2\tilde{J} +\tilde{D}-4K)}$ & \\\cline{4-5}
&  & $ H_{2} \leq H \leq\tilde{H}_{a}^\text{flop}$ & $\tilde{H}_{a}^\text{flop}=2/\tilde{M}%
\sqrt{(\tilde{D}+4K)(2\tilde{J}-\tilde{D}-4K)}$ & b-twin\\\cline{1-3}%
\multirow{2}{*}{$\tilde{E}_a^\text{min}-\tilde{E}_b^\text{min}$} &
\multirow{2}{*}{$=\frac{\tilde{M}^2 H^2}{2(2\tilde{J}+\tilde{D}-4K)} - \frac{\tilde{M}^2 H^2}{2(
2\tilde{J}-\tilde{D}-4K)} +2\tilde{D}$} &
$\tilde{H}_{a}^\text{flop}\leq H \leq H_{3}$ & $H_3=2/\tilde{M}%
\sqrt{2(\tilde{J}-2K)^{2}-\tilde{D}^{2}/2}$ & ($dE >0$)\\\cline{4-5}
&  & $H_{3}\leq H \leq\tilde{H}_{a}^\text{sat}$ & $\tilde{H}_{a}^\text{sat}=2/\tilde{M}%
(2\tilde{J}-\tilde{D}-4K)$ & a-twin\\\cline{1-3}%
$\tilde{E}_{a}^\text{sat}-\tilde{E}_{b}^\text{min}$ & $=-2\tilde{M} H+\frac{\tilde{M}^{2} H^{2}%
}{2(2\tilde{J}+\tilde{D}-4K)}-8K+4\tilde{J}$ & $\tilde{H}_{a}^\text{sat}\leq H \leq
\tilde{H}_{b}^\text{sat}$ & $\tilde{H}_{b}^\text{sat}=2/\tilde{M}(2\tilde{J}+\tilde{D}-4K)$ & ($dE <0$)\\
$\tilde{E}_{a}^\text{sat}-\tilde{E}_{b}^\text{sat}$ & $-2\tilde{D}$ &
\multicolumn{1}{l|}{$\tilde{H}_{b}^\text{sat}\leq H$} & \multicolumn{1}{c|}{-} & \\
\hline
\end{tabular}
\end{table*}

\subsection{Detwinning Dynamics} 

The full energy map for both types of domains in all fields allows us to calculate the energy cost/gain
of switching from a-twin domains to b-twin domains, $dE$, as a function of the reduced field, $MH/J_\perp$, and
reduced biquadratic coupling, $K/J_\perp$. While, 
 as  discussed in the main text, we have only
three experimentally observed phases, their dynamics depend on the energy difference $dE(H)$.

To illustrate this point, we show in Fig.~\ref{fig:E_ab2} the energy of the two types of domains in the small-field
regime for two specific instances of the $K/J_\perp$ ratio.
 While initially only the b-twin can reduce its energy, leading to the first phase, the energy
of the a-twin domain will start to reduce as well once $H_{\text{a}}^{\text{flop}}$ or $H_{\text{a}}%
^{\text{flip}}$ (depending on the ratio of $K/J_{\perp})$ is reached. Due to the minus sign in
the denominator of Eq.~(10) of the main text, this reduction is more pronounced
than in the b-twin domain, leading to a crossing of the two energies. At this
particular point, a domain reorientation from $b\parallel H $ to $a\parallel
H$ will be favored. Two cases are evident: if $12K/J_{\perp}<1$, $E^\text{min}_{b}$ will
intersect with $E^\text{min}_{a}$ at $H_{1}$, while for $12K/J_{\perp}>1$,
$E^\text{min}_{b}$ crosses with $E_{a}^{\text{sat}}$ at $H_{1}^{\text{sat}}>H_a^\text{sat}$.
where 
\begin{equation}
H_\text{1}^\text{sat}=\frac{4}{M}(J_\perp+4K)\left( 1-\sqrt{\frac{4K}{J_{\perp}%
+4K}}\right) ,\label{eq:H_ab}
\end{equation}

The difference in energy at high fields between $E_{b}^{\text{sat}}$ and
$E_{a}^{\text{sat}}$ is set by the biquadratic coupling $8K$. Once the the canting of the Fe moments in the b-twin
domain is significant, the energy of the b-twin will start to decrease,
 leading to the reorientation ($E_a^\text{sat}=\tilde{E}^\text{min}_b$) back to the initial phase, i.e. with $b\parallel H$. The corresponding field is given by 
\[
{H}_{2}=\frac{4}{\tilde{M}}\sqrt{K(2\tilde{J}+\tilde{D}-4K)}\approx \frac{4}{\tilde{M}}\sqrt{2K\tilde{J}}.
\] 

At extremely high magnetic fields the Fe moments
can spin-flop, which would lead to a final domain reorientation ($\tilde{E}_b^\text{min} = \tilde{E}_a^\text{min}$) with
$a\parallel H$. The associated field, however, is unphysically large:
\[
{H}_{3}=\frac{1}{\tilde{M}}\sqrt{8(\tilde{J}-2K)^2-2\tilde{D}^2}\approx\frac{2}{\tilde{M}}\sqrt{2}\tilde{J}>1100\,\text{T}.
\]

\begin{figure}
\includegraphics[width=0.5\textwidth]{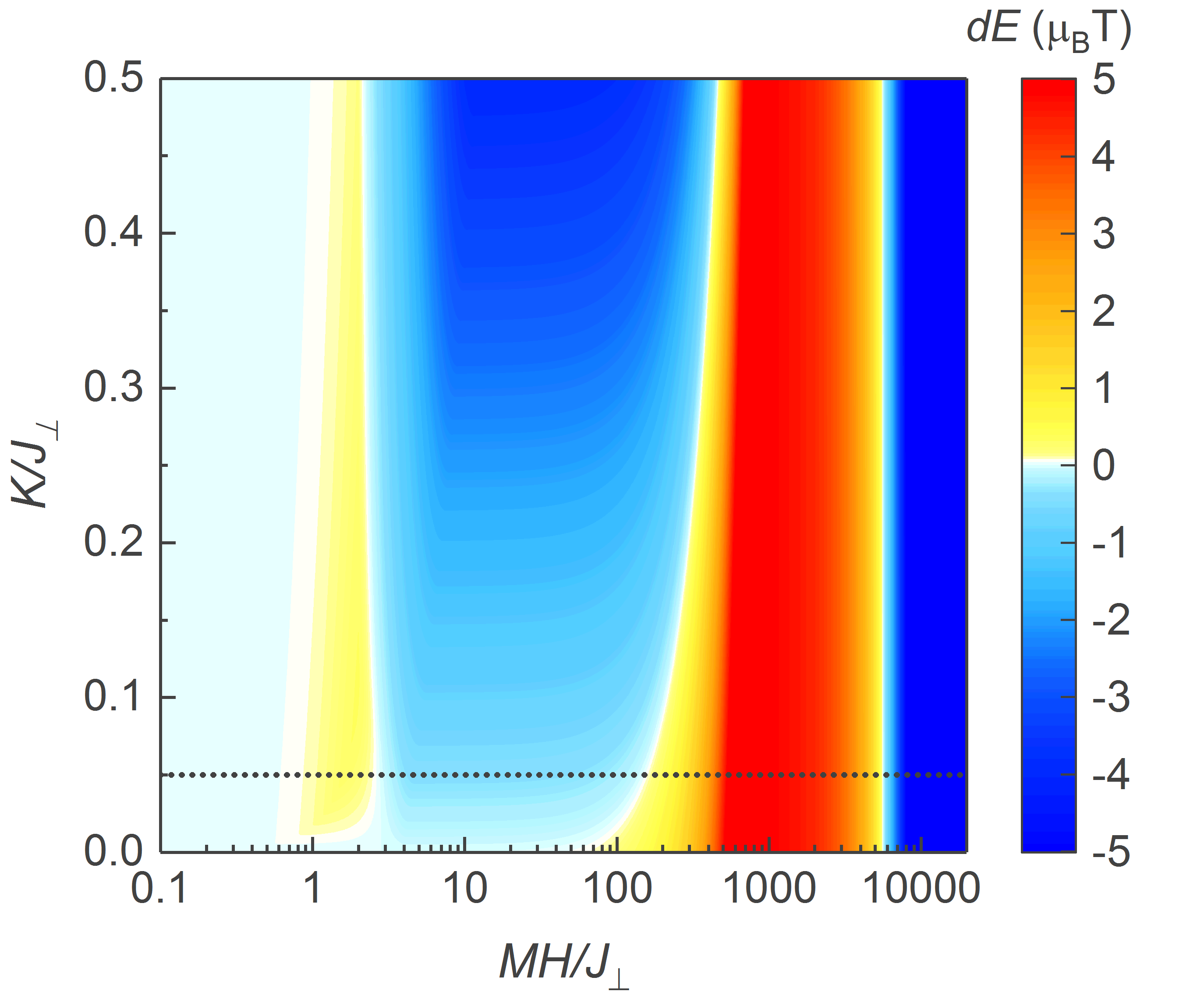}\caption{Calculated
domain energy difference $dE$ as a function of $K/J_{\perp}$ and $M
H/J_{\perp}$. The dotted line represents the cross section for $K/J_{\perp} =0.05$ shown in Fig.~(3) of the main text.}%
\label{fig:dE}%
\end{figure}

Since the domain dynamics are driven by the energy difference $dE$ between 
the domains we show in Fig.~\ref{fig:dE} $dE=E_a-E_b$ on a semi-log color map as 
a function of $K/J_{\perp}$ and $MH/J_{\perp}$. The dotted line represents $K/J_{\perp}=0.05$, for which 
$dE$ was depicted in the main text (Fig.~3). Positive and negative values 
correspond to the b-twin and a-twin domains being the ground state.  
All fields and regimes are summarized in Table~\ref{tab:summary} for the case $12K/J_\perp <1$.

\section{Determining the constants from another set of measurements}

\begin{table}
\caption{Coupling constants determined from various samples and experiments.
Equal superscripts in the Flux Method column indicate samples from the same batch. The corresponding data (MR, MS) in rows 2 and 3 are already published\cite{Jannis}. Other entries correspond to unpublished measurements.}%
\label{tab:results}%
\begin{tabular}
[c]{cccccc}%
Sample & Measurement & Flux & \multicolumn{3}{c}{Constants [meV]}\\
\#& Technique & Method & $J_{\perp}$ & $8K$ & $J_{\perp}/K$\\\toprule
1 & $M(H)$ & Fe-As$^{1}$ & 0.096 & 0.059 & 16\\
1 & MR & Fe-As$^{1}$ & 0.091 & 0.037 & 20\\
2 & MS & Bridgman$^{2}$ & 0.157 & 0.061 & 20\\
3 & $M(H)$ & Sn-flux$^{3}$ & 0.114 & 0.049 & 19\\
4 & Neutron & Sn-flux$^{3}$ & 0.116 & 0.062 & 15\\
5 & MR & Bridgman$^4$ & 0.130 & 0.092 & 11\\
6 & MR & Fe-As$^{5}$ & 0.142 & 0.053 & 22\\\hline
\end{tabular}
\end{table}

From the determination of the energy difference and the domain population the observed minimum (maximum) in magnetostriction, magneto-transport or field dependent neutron data around 0.4\thinspace T to 0.6\thinspace T
can be associated with $H_{a}^\text{flop}$. As Fig. 5 in the main text illustrates, this field, at which the energy difference
$|dE|$ reaches its first maximum (in the regime 
we are interested in) is close to the field $H_0$, where the population of the b-domain starts decreasing.
Thus, we can associate  $H_{a}^\text{flop}$ with $H_0$ and determine the relevant constants in a different way compared to that 
used in the main text:
\begin{align*}
J_{\perp}  & =\frac{M}{4}\frac{2H_{1}^{2}-H_{0}^{2}}{\sqrt{2(H_{1}%
^{2}-H_{0}^{2})}}\\
K  & =\frac{M}{16}\frac{H_{0}^{2}}{\sqrt{2(H_{1}^{2}-H_{0}^{2})}}%
\end{align*}
A check on the magnetoresistance data from the sample used in Zapf et
al.\cite{Jannis}, which shows $H_{0}=0.37\,\text{T}$ and $H_{1}%
=0.64$\thinspace T (both corrected for the observed hysteresis), yields
$J_{\perp}=0.091$\thinspace meV and $K=4.57$\thinspace $\mu$eV, which are in
agreement with the determination presented in the main text. See also Tab.~\ref{tab:results} (Sample 1). 
We also analyzed further samples grown with various synthesis techniques. The results are
 shown in Table~\ref{tab:results} as well.

\section{Extracting the twin population from magneto-transport data}

\begin{figure}
\includegraphics[width=0.5\textwidth]{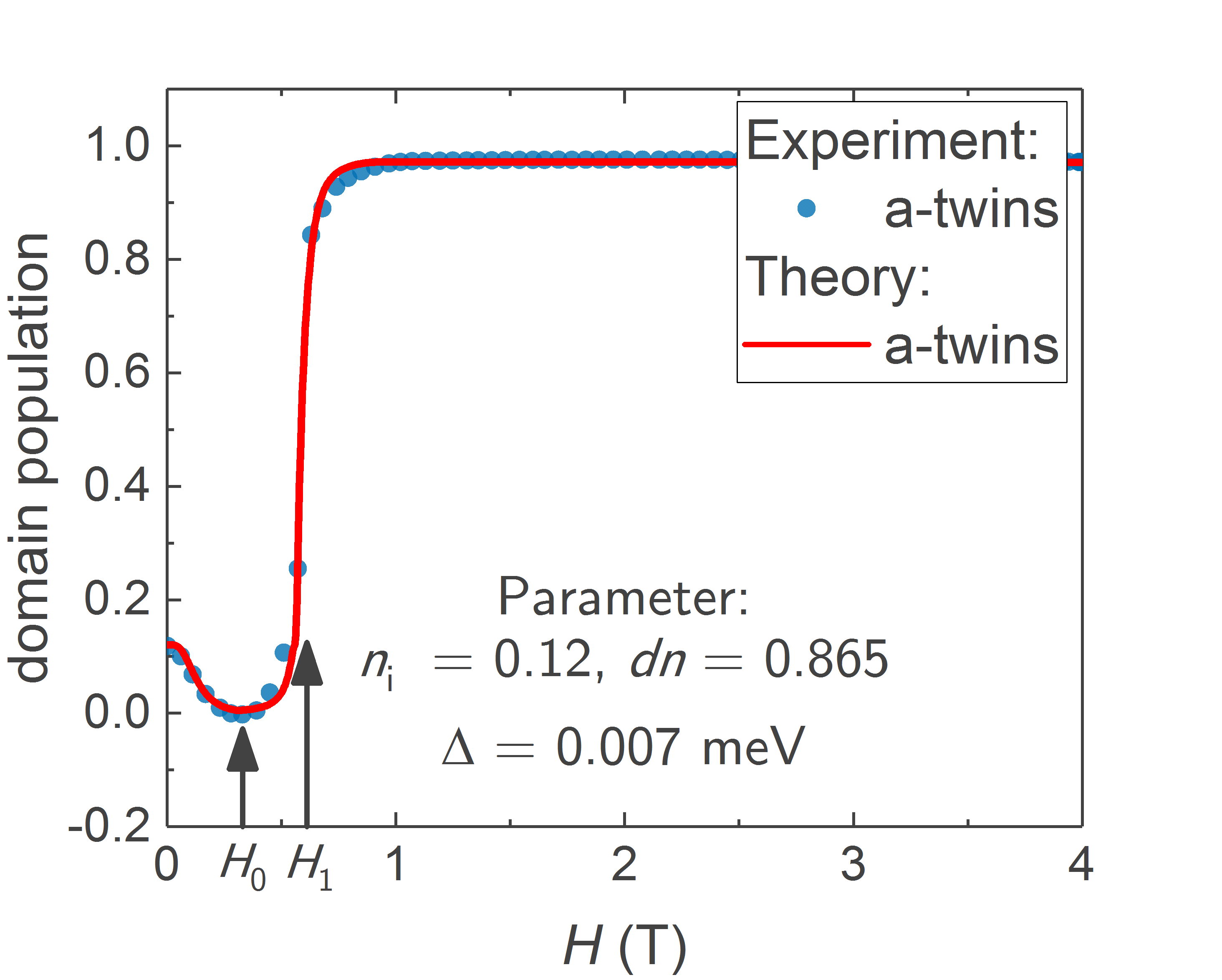}\caption{Twin domain population derived from magneto-transport\cite{Jannis} at $T=5$\,K as a function of (an decreasing) magnetic field applied along the [110]$_\text{T}$ direction (blue symbols). The solid (red) line presents the theoretical prediction for the given parameters.}%
\label{fig:MR}%
\end{figure}

Because the intrinsic resistivity of individual domains is anisotropic, and the domain population
depends on the magnetic field, we can extract the latter from the measured 
magnetoresistance (MR). In principle, the relation between MR and the twin population is nonlinear and requires 
solving the percolation problem. But since the intrinsic resistivity anisotropy is relatively small, 
this relation, whatever it is, can be linearized as if individual domains were connected in series. With this in mind,
and using the fact that our 
neutron data (which will be published separately) show the domain ratio to saturate at high fields with $n=0.976$,
we can simply rescale the MR from Ref.~\cite{Jannis} as 
\[
n(H)=\left(  1-\frac{R(H)-R_\text{sat}}{R(H_{0})-R_\text{sat}}\right)
\cdot n,
\]
(since the current
was perpendicular to the field direction in this experiment). Here
$R_\text{sat}$ denotes the MR value in the saturated state above $H_1$, corresponding to $n\approx1$. 
Figure~\ref{fig:MR} shows the result together with a fit to the obtained data.


%